\documentclass[longauth]{aa} % for the long lists of affiliations 
\usepackage{graphicx}
\usepackage{txfonts}
\usepackage{hyperref}
\usepackage[usenames,dvipsnames]{xcolor}
\hypersetup{colorlinks=true, citecolor=Turquoise, linkcolor=Turquoise, urlcolor=Turquoise}
\usepackage{amsmath}	% Advanced maths commands
\usepackage{amssymb}	% Extra maths symbols
\usepackage{epstopdf} % Convierte eps en pdf
\usepackage[normal]{threeparttable} %para poner notas en un pie de tabla
\usepackage{pstricks}
\usepackage{float}
\usepackage{subcaption}
\usepackage{multirow}
\usepackage[normalem]{ulem}
\usepackage{appendix}
\begin{document}

   \title{The multichord stellar occultation by the centaur Bienor on January 11, 2019}

   \subtitle{}

   \author{E. Fern\'andez-Valenzuela
          \inst{1,2}
\and
N. Morales\inst{2}%\fnmsep%\thanks{Just to show the usage of the elements in the author field}
\and{M. Vara-Lubiano\inst{2}}
\and{J. L. Ortiz\inst{2}}
\and{G. Benedetti-Rossi\inst{3,4,5}}
\and{B. Sicardy\inst{5}}
\and{M. Kretlow\inst{2}}
\and{P. Santos-Sanz\inst{2}}
\and{B. Morgado\inst{7,5,4}}
\and{D. Souami\inst{5,6}}
%Observadores
\and{F. Organero\inst{8}}
\and{L. Ana\inst{8}}
\and{F. Fonseca\inst{8}}
\and{A. Román\inst{9}}%Sociedad Astronomica Granadina
\and{S. Alonso\inst{10}}%Universidad de Granada
\and{R. Gonçalves\inst{11}}
\and{M. Ferreira\inst{12}}
\and{R. Iglesias-Marzoa\inst{13}}
\and{J.~L. Lamadrid\inst{13}}
\and{A. Alvarez-Candal\inst{2,14,7}}
\and{M. Assafin\inst{15,4}}
\and{F. Braga-Ribas\inst{16,7,4}}
\and{J.~I.~B. Camargo\inst{7,4}}
\and{F. Colas\inst{3}}
\and{J. Desmars\inst{17,18}}
\and{R. Duffard\inst{2}}
\and{J. Lecacheux\inst{5}}
\and{A.~R. Gomes-Júnior\inst{3,4}}
\and{F.~L. Rommel\inst{7,4}}
\and{R. Vieira-Martins\inst{7,4,5}}
\and{C.~L. Pereira\inst{7,4}}
%negatives
\and{V. Casanova\inst{2}}
\and{A. Selva\inst{19,20}}
\and{C. Perelló\inst{19,20}}
\and{S. Mottola\inst{21}}
\and{S. Hellmich\inst{21}}
\and{J.~L. Maestre\inst{22}}
\and{A. J. Castro-Tirado\inst{2}}
\and{A. Pal\inst{23}}
\and{J.~M. Trigo-Rodriguez\inst{24,25}}
\and{W. Beisker\inst{20}}
\and{A. Laporta\inst{26,27}}
\and{M. Garcés\inst{26,27}}
\and{L. Escaned\inst{26,27}}
\and{M. Bretton\inst{28}}
          }

   \institute{Florida Space Institute, UCF, 12354 Research Parkway, Partnership 1 building, Room 211, Orlado, USA. \\ \email{estela@ucf.edu}%1
\and
 Instituto de astrof\'isica de Andaluc\'ia, CSIC,
              Glorieta de la Astronom\'ia s/n, 18008 Granada, Spain%2
\and
UNESP - S\~ao Paulo State University, Grupo de Din\^amica Orbital e Planetologia, Guaratinguet\'a, SP, 12516-410, Brazil%3
\and
Laborat\'orio Interinstitucional de e-Astronomia - LIneA - and INCT do e-Universo, Rua Gal. Jos\'e Cristino 77, Rio de Janeiro, RJ, 20921-400, Brazil%4
\and
LESIA, Observatoire de Paris, Universit\'e PSL, Sorbonne Universit\'e, Universit\'e de Paris, CNRS, 92190 Meudon, France%5
\and
naXys, University of Namur, 8 Rempart de la Vierge, Namur, B-5000, Belgium%6
\and
Observat\'orio Nacional (MCTI), Rua Gal. Jos\'e Cristino, 77—Bairro Imperial de S\~ao Crist\'ov\~ao, 20921-400 Rio de Janeiro, Brazil%6'
\and
Complejo Astron\'omico de La Hita, Camino de Do\~na Sol, s/n, 45850 LaVilla de Don fadrique, Spain%7
\and
Sociedad Astron\'omica Granadina, Apartado de Correos 195, 18080 Granada, Spain%8
\and
Software Engineering Department, Andalusian Research Institute in Data Science and Computational Intelligence, University of Granada, Spain%9
\and
Instituto Polit\'ecnico de Tomar; CI2 e U.D. Matem\'atica e F\'isica, 2300-313 Tomar, Portugal%10
\and
Centro Ci\^encia Viva, Alto de Santa B\'arbara, Via Galileu Galilei 817, 2250-100 Const\^ancia, Portugal%11
\and
Centro de Estudios de F\'isica del Cosmos de Arag\'on, Plaza San Juan 1, 44001 Teruel, Spain%12
\and
Instituto de F\'isica Aplicada a las Ciencias y las Tecnolog\'ias, Universidad de Alicante, San Vicent del Raspeig, E03080, Alicante, Spain%13
\and
Observat\'orio do Valongo/UFRJ, Ladeira Pedro Antonio 43, Rio de Janeiro, RJ 20.080-090, Brazil%15
\and
Federal University of Technology-Paraná (UTFPR/DAFIS), Av. Sete de Setembro, 3165, CEP 80230-901 - Curitiba - PR - Brazil%16
\and
Institut Polytechnique des Sciences Avancées IPSA, 63 boulevard de Brandebourg, F-94200 Ivry-sur-Seine, France%17
\and
Institut de Mécanique Céleste et de Calcul des Éphémérides, IMCCE, Observatoire de Paris, PSL Research University, CNRS,Sorbonne Universités, UPMC Univ Paris 06, Univ. Lille, 77 Av. Denfert-Rochereau, F-75014 Paris, France%18
\and
Agrupaci\'on Astron\'omica de Sabadell, Carrer Prat de la Riba, 116, 08206 Sabadell, Spain%19
\and
International Occultation Timing Association / European Section, Am Brombeerhag 13, 30459 Hannover, Germany%20
\and
Institute of Planetary Research, DLR Rutherfordstr. 2, D-12489 Berlin, Germany%21
\and 
Albox Observatory, Almer\'ia, Spain%22
\and
Konkoly Observatory, Research Centre for Astronomy and Earth Sciences, H-1121 Budapest, Konkoly Thege Miklós út 15-17, Hungary%23
\and
Institute of Space Sciences (ICE, CSIC), Carrer de Can Magrans, s/n, Campus UAB,
08193 Cerdanyola del Vall\'es (Barcelona), Catalonia, Spain.%24
\and
Institut d'Estudis Espacials de Catalunya (IEEC), Edif. Nexus, c/Gran Capit\`a, 2-4, 08034 Barcelona, Catalonia, Spain%25
\and
Agrupaci\'on Astron\'omica de Huesca. Parque Tecnol\'ogico Walqa, 13. 22197 Cuarte, Huesca, Spain%26
\and
Observatorio Astron\'omico Torres de Alcanadre, 22132 Torres de Alcanadre, Spain%27
\and
Observatoire des Baronnies Proven\c{c}ales, Le Mas des Gr\`es, Route de Nyons, 05150 Moydans, France%28
}

 \date{Received XX XXX, XXXX; accepted XX XX, XXXX}

\abstract{Within our program of physical characterization of trans-Neptunian objects and centaurs, we predicted a stellar occultation by the centaur (54598) Bienor to occur on January 11, 2019, with good observability potential. We obtained high accuracy astrometric data to refine the prediction, resulting in a shadow path favorable for the Iberian Peninsula. This encouraged us to carry out an occultation observation campaign that resulted in five positive detections from four observing sites. This is the fourth centaur for which a multichord (more than two chords) stellar occultation has been observed so far, the other three being (2060) Chiron, (10199) Chariklo, and (95626) 2002 GZ$_{32}$. From the analysis of the occultation chords, combined with the rotational light curve obtained shortly after the occultation, we determined that Bienor has an area-equivalent diameter of $150\pm20$ km. This diameter is $\sim30$ km smaller than the one obtained from thermal measurements. The position angle of the short axis of the best fitting ellipse obtained through the analysis of the stellar occultation does not match that of the spin axis derived from long-term photometric models. We also detected a strong irregularity in one of the minima of the rotational light curve that is present no matter the aspect angle at which the observations were done. We present different scenarios to reconcile the results from the different techniques. We did not detect secondary drops related to potential rings or satellites. Nonetheless, similar rings in size to that of Chariklo's cannot be discarded due to low data accuracy.}


 \maketitle
%
%-------------------------------------------------------------------
\section{Introduction}
\label{sec:introduction}

Centaurs are objects that orbit the Sun with orbital semi-major axes between those of Jupiter and Neptune. Dynamical evolution models show that these objects originated in the trans-Neptunian region, but they have been injected into inner parts of the Solar System as a result of planetary encounters, mostly with Neptune. Dynamical simulations also estimate that the mean half-life of the entire sample of 32 well-known centaurs is 2.7 Myr \citep{Horner2004a}, which means they have spent most of their lifetime in the trans-Neptunian region. Therefore, centaurs present an excellent opportunity to study smaller trans-Neptunian objects (TNOs) much closer to the Earth, providing a better characterization of their physical properties. Due to their short lifetime in their current region, centaurs have suffered less chemical or radiolytic processes than other bodies (e.g., asteroids and comets). Hence, the study of the physical properties of this population 
provides important information to help reconstruct the history of the Solar System formation and evolution.

During the last years, there has been a growing interest in the centaur population due to the discovery of a ring system around (10199) Chariklo \citep{Braga-Ribas2014}. Also, the data of several stellar occultations by (2060) Chiron reveal features similar to those found for Chariklo; although, there is still an open debate on whether these features are attributed to rings \citep{Ortiz2015,Ruprecht2015,Sickafoose2020}. The discovery of ring systems has opened a new branch of research to understand how these rings are formed and how they survive around these small bodies. Collisions and rotational disruptions seem to be plausible scenarios for their formation \citep[see,][and references therein]{Braga-Ribas2014,Ortiz2015,Sicardy2019}. Additionally, the discovery of another ring system around the dwarf planet Haumea \citep{Ortiz2017} raises the question of whether rings are formed in the trans-Neptunian region surviving through the planetary encounters \citep[which does not seem to be very unlikely,][]{Araujo2016}. Another proposed scenario is by tidal disruption of a satellite when encountering the giant planets \citep{Hyodo2016}; although, the probability of such an event seems to be very small \cite[e.g.,][]{Melita2017}. Moreover, this would not explain Haumea's system \citep[which includes the only known dynamical family of objects in the trans-Neptunian region,][]{Brown2007,Ortiz2012b,Proudfoot2019,Ortiz2020abstract}. Regardless, these three objects, Chariklo, Chiron, and Haumea, share some physical properties; for instance, they all have highly elongated ellipsoidal shapes, rotate relatively fast, and present water ice in their unresolved spectra, \cite[i.e., main body plus rings, see, e.g.,][and references therein]{Barucci2011a,Braga-Ribas2014,Duffard2014,Ortiz2017,Cikota2018,Sicardy2019,Morgado2021}, and most likely those properties are related to the existence of rings around them.

(54598) Bienor, with the provisional designation 2000~QC$_{243}$, is one of the largest centaurs known to date and slightly smaller than that of Chiron and Chariklo \citep{Braga-Ribas2014,Ortiz2015}. Its size was determined using radiometric measurements from Herschel Space Observatory (Herschel) by \cite{Duffard2014b} where a diameter of $198^{+6}_{-7}$ km is given. Later, \cite{Lellouch2017} updated Bienor's size to be in the range of $179 - 184\pm 6$ km (depending on the model and geometry after including ALMA's data)\footnote{For future references within this work, we use the average measurement (with its corresponding uncertainty) from the range given by \cite{Lellouch2017}, which results in $181\pm8$ km.}. Bienor also shares its ellipsoidal shape and a watery spectrum with these two objects \cite[e.g.,][]{Dotto2003,Rabinowitz2007,Fernandez-Valenzuela2017}; although, its rotation period ($\sim9.14$ h) is slightly over the average of the TNO and centaur population \citep{Thirouin2014}. These features led \cite{Fernandez-Valenzuela2017} to propose that Bienor could possess a ring system, similar to those found in the abovementioned small bodies. In fact, Bienor's photometric data along the years present a steep increase in brightness of around 0.7 mag and a decrease in its rotational light curve amplitude of around 0.6 mag, which seems to be consistent with a ring model. However, other possibilities also considered in \cite{Fernandez-Valenzuela2017} fit considerably well, for instance, a highly elongated body out of hydrostatic equilibrium.

Within our international collaboration to detect stellar occultations by outer Solar System bodies, we predicted that Bienor would occult the Gaia DR2 236278279941456256 star on January 11$^{th}$, 2019. This presented a great opportunity to shed light on the reason behind the sharp increase in Bienor's brightness. The initial prediction was done using the numerical integration of the motion of an asteroid (NIMA) solution \citep{Desmars2015a,Desmars2015b}. We refined this prediction performing specific observations as described in Section \ref{sec:prediction}. The stellar occultation was successfully observed from four different sites (Section \ref{sec:observations}), with four other sites reporting no star occultation. We present the data reduction and results in Section \ref{sec:data_reduction}. Additionally, we obtained Bienor's rotational light curve in order to help improve our interpretation of the stellar occultation (see Section \ref{sec:RLC_observations}). Finally, we provide a thorough analysis in Section \ref{sec:analysis}, the discussion of our results in Sections \ref{sec:Discussion}, and our final conclusions in Section \ref{sec:conclusions}.

%--------------------------------------------------------------------------------------------------------------

%--------------------------------------------Observations-------------------------------------------------------

\section{Observations for predicting Bienor's stellar occultation}
\label{sec:prediction}

In order to update and refine the occultation prediction for Bienor on January 11, 2019, we took images of the centaur on December 6, 2018 and January 2, 2019 (closer to the occultation event). The observations were carried out with the 1.23-m Calar Alto Observatory telescope in Almer\'ia (Spain). We used the 4k$\times$4k DLR-MKIII CCD camera, which has an pixel scale and a field of view (FoV) of $0.32''$pixel$^{-1}$ and $21'\times21'$, respectively. Images were taken using $2\times2$ binning mode, the $R$-Johnson filter, and 300 s exposure time. The average seeing of each night was $1.4''$ on December 6, 2018, and $1.3''$ on January 2, 2019. We experienced good weather and dark nights under new moon and $16\%$ of moon brightness during the first and second runs, respectively.

\begin{figure}
    \includegraphics[width=\columnwidth]{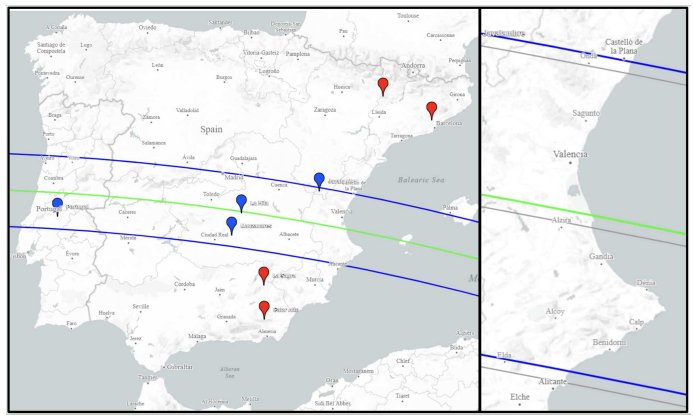}

    \caption{
    Map representing the post-occultation shadow path (blue lines) of the stellar occultation produced by Bienor on 11$^{th}$ January 2019 (with the green line representing the center of the shadow). The width of the shadow path on the map is that from the fitted parameters of S1 scenario (see Section \ref{sec:analysis}). The motion of the shadow is from East to West. Blue and red tags show the location of the observatories in Table \ref{tab:observations_StellarOccultation}, with blue meaning positive detection and red meaning negative detection. At the right it can be seen a zoom of the Spanish East coast displaying the difference between the predicted (gray lines) and the post-occultation shadow path (with the same colour code as in the left panel), which results into 10 mas. }
   
    \label{fig:path}
\end{figure}

In both runs, bias frames and twilight sky flat-field frames were taken at the beginning of each observation night. A median bias was subtracted from all images (flat field and science frames). A normalized flat-field (after median bias subtraction) was obtained and used to correct the science images. The signal-to-noise ratio (S/N) for Bienor was $\sim$ 60 and 63 for the first and second runs, respectively. We carried out astrometric measurements using our own codes written in  {\ttfamily Python} that select the different stars within the images and compare them with the Gaia DR2 catalog \citep{Gaia-Collaboration2018}, obtaining the astrometric position of the target related to the stars. This astrometric position is then used to update the prediction of the stellar occultation by using the methodology explained in \cite{Ortiz2020}. We obtained astrometric data with a precision of 9 milliarcseconds (mas) in both runs, which includes the plate solution and the centroid uncertainty. Considering the JPL$\#$45 orbit, the observation astrometric offset with respect to the initial prediction in R.A.$\times\cos({\rm Dec.})$ and Dec. was +127 mas and -17 mas, respectively. The path predicted for the occultation after updating the astrometric position of Bienor is represented in Figure \ref{fig:path} (right panel) by gray lines.

\section{Observations of the stellar occultation}
\label{sec:observations}

\begin{table*}
	\centering
	\caption{Information about the occulted star.}
	\label{tab:star_information}
	\begin{tabular}{cccc}   
		%\hline	
		\hline	    
		\textbf{Designation}   &  \textbf{Coordinates}$^{a}$    &  \textbf{Magnitudes}$^{b}$ &\textbf{Diameter}$^{c}$     \\
\hline
Gaia DR2 236278279941456256 & 	$\alpha= 03{\rm h}\, 19{\rm m}\, 57{\rm s}.7756$        & $G=16.00157$	 &$\sim 0.0055$ mas\\
UCAC4 654-014368            & $\delta= +40^\circ\, 40'\, 53''.2215$ &  $V=16.329$; $K= 14.417$
 & 	 $\sim 0.056$ km at Bienor's distance \\	
\hline
	\end{tabular}
\begin{tablenotes}
      \small
  	  \item $^{a}$Gaia DR2 coordinates corrected from proper motion for the occultation date.
      \item $^{b}$From the Gaia ($G$) and UCAC4 \citep[$V$, $K$;][]{Zacharias2012} catalogs.
	  \item $^{c}$Estimated from the V, K magnitudes of the star, using equation \eqref{eq:vanBelle}. For a detailed information see Section \ref{sec:analysis}.
\end{tablenotes}
\end{table*}

On January 11, Bienor occulted the star Gaia DR2 236278279941456256 (or UCAC4 654-014368, see Table \ref{tab:star_information} for detailed information regarding the occulted star). A total of nine different telescopes at eight different locations acquired data to support this stellar occultation (see Figure \ref{fig:path}). Table \ref{tab:observations_StellarOccultation} presents a list with the names of the observing sites involved in this event, their geodetic coordinates (longitude, latitude, and altitude), diameters and focal distance of the telescopes, detector manufacturers and models, observer names, exposure time ($T_{\rm exp}$), cycle time (total time between consecutive exposures) for the observations, and status of the detection. A total of five positive and three negative detections were obtained as detailed in the table. 

We obtained time-series observations using each of the telescopes (and cameras) listed in Table \ref{tab:observations_StellarOccultation}. All computers were synchronized using Network Time Protocol (NTP) servers or Video Time Inserters which are time-synchronized by Global Positioning Systems (GPS) devices. The acquisition time of each image was inserted on the corresponding image header. All time-series of CCD images were started from around 15 minutes before and ended 15 min after the predicted occultation time in order to obtain both a good base-line to characterize the occultation and the noise level before and after the event. No filters were used in any of the observatories to maximize the S/N of the occulted star (blended with Bienor), obtaining the best possible photometry.

\begin{table*}
	\centering
	\caption{Observatory, telescope, and observation details for the stellar occultation.} 
	\label{tab:observations_StellarOccultation}
	\begin{tabular}{lccccc} 
		\hline\hline
\textbf{Observatory} 	& \textbf{Coordinates} & \textbf{Telescope details}  & \textbf{Observers}& \textbf{Observation details}& \textbf{Detection}\\
\hline            

Name	& Longitude (W)  & Diameter           & & $T_{\rm exp}$ (s)                        & \\
City    &  Latitude (N)  & Focal distance     &  &Cycle time (s)	       &\\
Country & Altitude (m)   & Detector/Instrument &  &	        Synchronization               &\\
 \hline
 \hline
Javalambre$^{\dag}$      & $01^\circ\, 00'\, 58.58''$  & D = 28.0 cm 	           & R. Iglesias & 4.0               		 & 		   \\
Teruel          & $40^\circ\, 02'\, 30.58''$  & 	f = 1764 mm 			& J.L. Lamadrid & 6.928  & Positive\\
Spain		    & 1957		              & 	SBIG ST10XME		& 		  & NTP  & 			\\

 \hline
La Hita         & $03^\circ\, 11'\, 09.7''$   & D = 77 cm 			    & N. Morales	    & 2.0    	  & 		   \\
Toledo          & $39^\circ\, 34'\, 06.8'' $  & 	f = 2656 mm 	 	& F. Organero		& 2.957       & Positive\\
Spain		    & 695           		  & FLI PROLine 16803		& 		            & 	NTP& 			\\				    
 \hline
La Hita         & $03^\circ\, 11'\, 09.7''$   & D = 40 cm  				& F. Organero       & 4.0   	   & 		   \\
Toledo          & $39^\circ\, 34'\, 06.8''$   & 	f = 1800 mm			& L. Ana     	    & 4.0	       & Positive\\
Spain	        & 695           		  & Basler acA640-120 $\mu$m& F. Fonseca	    & 	NTP & 			\\	
 \hline
Manzanares      & $03^\circ\, 27'\, 23.3''$   & D = 60 cm  			    & S. Alonso		    & 1.495    	   & 		   \\
Ciudad Real     & $39^\circ\, 04'\, 55.2''$   & f = 3125 mm				& A. Rom\'an  	    & 1.495        & Positive\\
 Spain			& 132           		  & ZWO ASI174MM-Cool		& 		            & NTP   & 			\\	
 \hline

Centro Ci\^encia Viva   & $08^\circ\, 19'\, 25.2''$	 & D = 50.8 cm,     & 	R. Gon\c{c}alves & 2.56 		 & 		    \\
Const\^ancia        	& $39^\circ\, 29'\, 41.6''$  & f = 3454 mm      & M. Ferreira  	 	 & 2.56          & Positive \\
Portugal      		    & 147           		 & WATEC 910HX-RC	& 		  	         & 	GPS	 & 			\\					    
\hline
\hline

Agrup. Astro. Sabadell$^\ddag$  & $357^\circ\, 54'\, 57.4''$  & D = 50 cm    & A.Selva		&  2.56 	&			\\
Sabadell				        & $41^\circ\, 33'\, 00.2$     & f = 2000 mm	 & C. Perell\'o & 2.56      & Negative \\    					
Spain					        & 224					  &WATEC 910HX-RC& 		        & GPS& 			\\
 \hline
La Sagra   		    & $02^\circ\, 33'\, 52''$   & D = 36 cm 		   & N. Morales &  3.0	 	 &			\\
Granada				& $37^\circ\, 59'\, 02''$   &  f = 720 mm		   &            &  3.0       & Negative \\   			   
Spain				& 1530				    & 			QHY174M-GPS& 		  	& GPS & 			\\	
\hline
Calar Alto          & $02^\circ\, 32'\, 46''$   & D = 123 cm            &   S. Mottola      & 0.978         &   \\
Almer\'ia           & $37^\circ\, 13'\, 25''$   & f = 9880 mm           & S. Hellmich       & 1.246         &    Negative \\
Spain               &   2168                &  DLR-MKIII            &                   &   NTP  &       \\
\hline
TJO$^{\star}$           & $0^\circ\, 43'\, 46.74''$ & D = 80 cm         &                       & 3 & \\
San Esteban de la Sarga & $42^\circ\, 3'\, 5.95''$  & f = 7680 mm       &                       & 6 & Negative \\
Spain                   &       1570  & MEIA2             &     D. Souamy                  &  NTP &\\
 \hline\hline

	\end{tabular}
	\begin{tablenotes}
	\item $^{\dag}$Excalibur telescope at Javalambre Astrophysical Observatory
     \item $^\ddag$Agrupaci\'on Astron\'omica de Sabadell
     \item $^\star$Telescopio Joan Or\'o
      
    \end{tablenotes}
\end{table*}

\section{Data reduction and results from the stellar occultation}
\label{sec:data_reduction}

Images were bias and flat-field calibrated using standard procedures as described in Section \ref{sec:prediction}. In the case of the 40-cm telescope at La Hita and Constancia, the observations were recorded in video format. We converted the video into FITS prior to the analysis using Tangra v3.6.18\footnote{http://www.hristopavlov.net/Tangra3/}. 

\begin{figure}
    \includegraphics[width=\columnwidth]{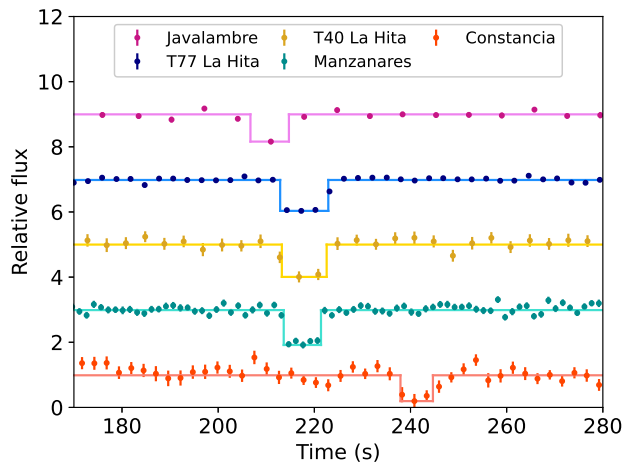}
    \caption{Occultation light curves with positive detections obtained from different telescopes distributed along the shadow path. colour points represent observational data with their uncertainties, while the lines represent a square-well fit to the observational data. Time is given in seconds after 2019 January 11 01:00:00 UT. Light curves have been normalized to one outside the occultation drop and shifted by an arbitrary offset in this plot for clarity. \label{fig:Occultation_LC}}
\end{figure}

We performed synthetic aperture photometry in order to obtain the flux of the occulted star blended with the centaur relative to a set of reference stars. We used our own routines written in IDL, which are based on \texttt{daophot} package \citep{Stetson1987}. These routines are specifically designed to measure the flux from the combination of the occulted star plus the small body. In contrast to other software routines, that obtain the centroid within the aperture to extract the flux, our routine places the centroid at a fixed position relative to the reference stars calculating the photo-center of the star prior to the occultation, thus avoiding errors due to the wandering of the photo-center when the star is occulted and the signal drops to nearly zero. We applied relative photometry using the stars present in the FoV in order to minimize systematic photometric errors due to atmospheric variability. As a result, we obtain the combined flux of the star and the object versus time, in other words, the light curve of the occultation for each telescope. As the event only lasts several seconds, flux variations due to rotational variability of the small body do not affect the resulting light curve. A total of five positive events and two close negative events were obtained. Other negatives farther away from the body were also obtained as detailed in Table \ref{tab:observations_StellarOccultation}. We plot the positive light curves in Figure \ref{fig:Occultation_LC}. While the positive chords are used to model the size and shape of the body, the negative chords limit the body limb providing useful constraints.

From the positive detections, the light curves are used to derive the time at which the star disappears (\textit{ingress}) and reappears (\textit{egress}) behind Bienor's limb. The total duration of the occultation for each light curve is therefore delimited by the ingress and egress timings, which translates into different segments (known more commonly as chords) in the plane of the sky for each observatory using Bienor's apparent motion (13.47 km/s at the time of the occultation). The set of different chords as a function of their location provides the shape and size of the body, as explained in Section \ref{sec:analysis}. The ingress and egress timings were obtained using \texttt{SORA} v0.1\footnote{\href{https://sora.readthedocs.io/overview.html}{Stellar Occultation Reduction Analysis (SORA).}} \citep{Gomes-Junior2022}, where each light-curve is fitted to a square-well function as described by \cite{Ortiz2017} and references therein. The uncertainties in the timing are obtained by evaluating the fit by a $\chi^2$ function as explained in the supplementary documentation of the \cite{Sicardy2011} and in \cite{Souami2020}. There are three main sources of uncertainties involved in the ingress and egress timing calculation (considering that the synchronization of the cameras was correct): the Fresnel diffraction, the angular diameter of the star, and the cycle times of the different detectors.

The projected diameter of the occulted star (Gaia DR2 236278279941456256) in the plane of the sky can be estimated using the equation:
\begin{equation}
\label{eq:vanBelle}
    \theta=\frac{\theta_{V=0}}{10^{m_V/5}},
\end{equation}
where $m_V$ is the apparent magnitude of the star in $V$-band, and $\theta_{V=0}$ is given by the equation:
\begin{equation}
    \theta_{V=0}=10^{(0.5+0.264(V-K))},
\end{equation}
\citep[for main sequence stars;][]{vanBelle1999}, where $(V-K)$ is the corresponding colour of the star. Using the star's $V$, and $K$ apparent magnitudes (see Table \ref{tab:star_information}), we obtained an angular diameter $\theta=0.0055$ mas, which translates into 0.056 km at Bienor's geocentric distance at the time of the occultation ($\Delta = 14.16$ au). This projected diameter is one order of magnitude smaller than the Fresnel scale, which is given by the equation
\begin{equation}
F=\sqrt{\lambda \Delta/2} = 0.797\, {\rm km},
\end{equation}
where $\lambda = 600$ nm is the average central wavelength for the CCD observations. While the errors from the shortest cycle time of detectors used in this event, which is 1.495~s, translates into 20 km (taking into account Bienor's velocity of 13.5 km$\,$s$^{-1}$ at the moment of the occultation). Therefore, the uncertainty introduced in the ingress and egress timing from Fresnel diffraction effects and/or from the angular diameter of the star are negligible when compared with those coming from the cycle time used in each detector and the error resulting from the data dispersion when fitting a square well. We note that this calculation does not account for uncertainties due to synchronization problems that might cause shifts in the chords. As discussed in Section \ref{sec:analysis} those errors become apparent when plotting the chords and are discussed in the analysis section. Table \ref{tb:ingress_egress} presents the results obtained for the ingress and egress timings from each light curve.

\begin{table*} 
	\centering
	\caption{Ingress and egress timings and chord lengths derived from the occultation light curves.
	}
	\label{tb:ingress_egress}
	\begin{tabular}{ccccccccc}   

		\hline	    
Observatory & Ingress          & Egress           & Chord length & $\sigma$  & Shift S1\tablefootmark{$^\dag$}    & Shift S1\tablefootmark{$^\dag$} & Shift S2\tablefootmark{$^\dag$}    & Shift S2\tablefootmark{$^\dag$}\\
  & (UT)  & (UT)    & (km)       &    &(km)     & (s)&(km)     & (s)\\
\hline\hline
Javalambre	& 01:03:27.8 $\pm$1.5      & 01:03:34.7 $\pm$1.5      & $96\pm29$	    & 0.08      & 0 & 0 & 35    & 2.6\\			
La Hita T77 & 01:03:33.2 $\pm$ 0.7  & 01:03:43.2 $\pm$ 0.2  & $139\pm10$	    & 0.07      & 0 & 0 & 14    & 1.01\\
La Hita T40	& 01:03:33.5 $\pm$ 0.5  & 01:03:42.8 $\pm$ 0.6  & $129\pm11$        & 0.18      & 0 & 0 & 14    & 1.01\\
Manzanares  & 01:03:33.9 $\pm$ 0.1  & 01:03:41.7 $\pm$ 0.2  & $107\pm3$         & 0.11      & -54.7 & -4.06 & -7 & -0.5\\
Constancia  & 01:03:58.2 $\pm$ 0.7  & 01:04:04.9 $\pm$ 0.5  & $94\pm12$         & 0.19      & 0 & 0 & 40    & 3.0\\			
		\hline
		%\hline
	\end{tabular}
	\tablefoot{Abbreviations are defined as follows: the dispersion of the light curve removing the points from the occultation ($\sigma$).\\ $^\dag${Shifts applied to the chords in scenario S1 (same shifts are applied in S3) and S2 (Section \ref{sec:analysis}). A positive shift indicates a displacement toward the southwest, whereas a negative shift indicates one toward the northeast.}}
\end{table*}

%%%%%%%%%%%%%%%%%%%%%%%%%%%%%%%%%%%%%%%%%%%%%%%%%%%%%%%%%%%%%%%%%%%%%%%%%%%%%%%%%%%%%%%%%%%%%%%%
%%%%%%%%%%%%%%%%%%%%%%%%%%%%%%%%%%%%%%%%%%%%%%%%%%%%%%%%%%%%%%%%%%%%%%%%%%%%%%%%%%%%%%%%%%%%%%%%
%% rotational light curve
%%%%%%%%%%%%%%%%%%%%%%%%%%%%%%%%%%%%%%%%%%%%%%%%%%%%%%%%%%%%%%%%%%%%%%%%%%%%%%%%%%%%%%%%%%%%%%%%
%%%%%%%%%%%%%%%%%%%%%%%%%%%%%%%%%%%%%%%%%%%%%%%%%%%%%%%%%%%%%%%%%%%%%%%%%%%%%%%%%%%%%%%%%%%%%%%%

\section{Rotational light curve: Observations, data reduction, and results}
\label{sec:RLC_observations}

In order to determine Bienor's rotational phase at the time of the stellar occultation, we carried out an observational campaign during the second semester of 2019 with the goal of obtaining its rotational light curve. We observed Bienor during seven different nights using two telescopes as described below.

On September 23 and 24, we used the 1.23-m telescope of Calar Alto Observatory, in Almer\'ia (Spain). This telescope has a 4k$\times$4k CCD with a FoV of $21.5'\times21.5'$ and a pixel scale of $0.315''/$pixel. Thanks to the large FoV, we were able to aim the telescope at the same coordinates during both nights. This is convenient as it allowed us to use the same set of stars to perform relative photometry, avoiding systematic errors associated with absolute photometric techniques. The average seeing in this run was $1.86''$.

\begin{figure*}
	\begin{subfigure}{.5\textwidth}
        \centering
        \includegraphics[width=\linewidth]{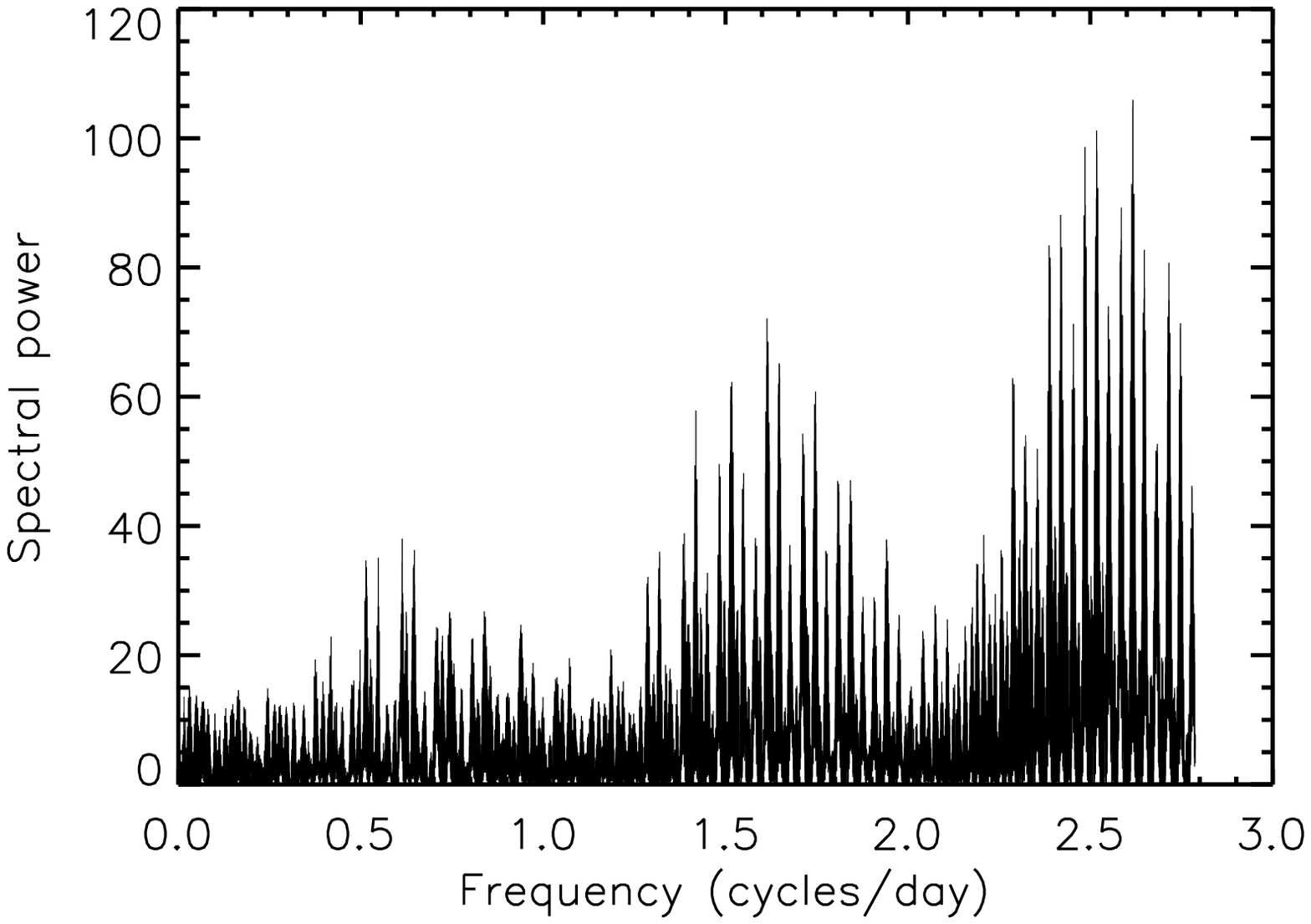}
        \caption{Lomb}
        \label{fig:Lomb}
    \end{subfigure}%
	    \begin{subfigure}{.5\textwidth}
        \centering
        \includegraphics[width=\linewidth]{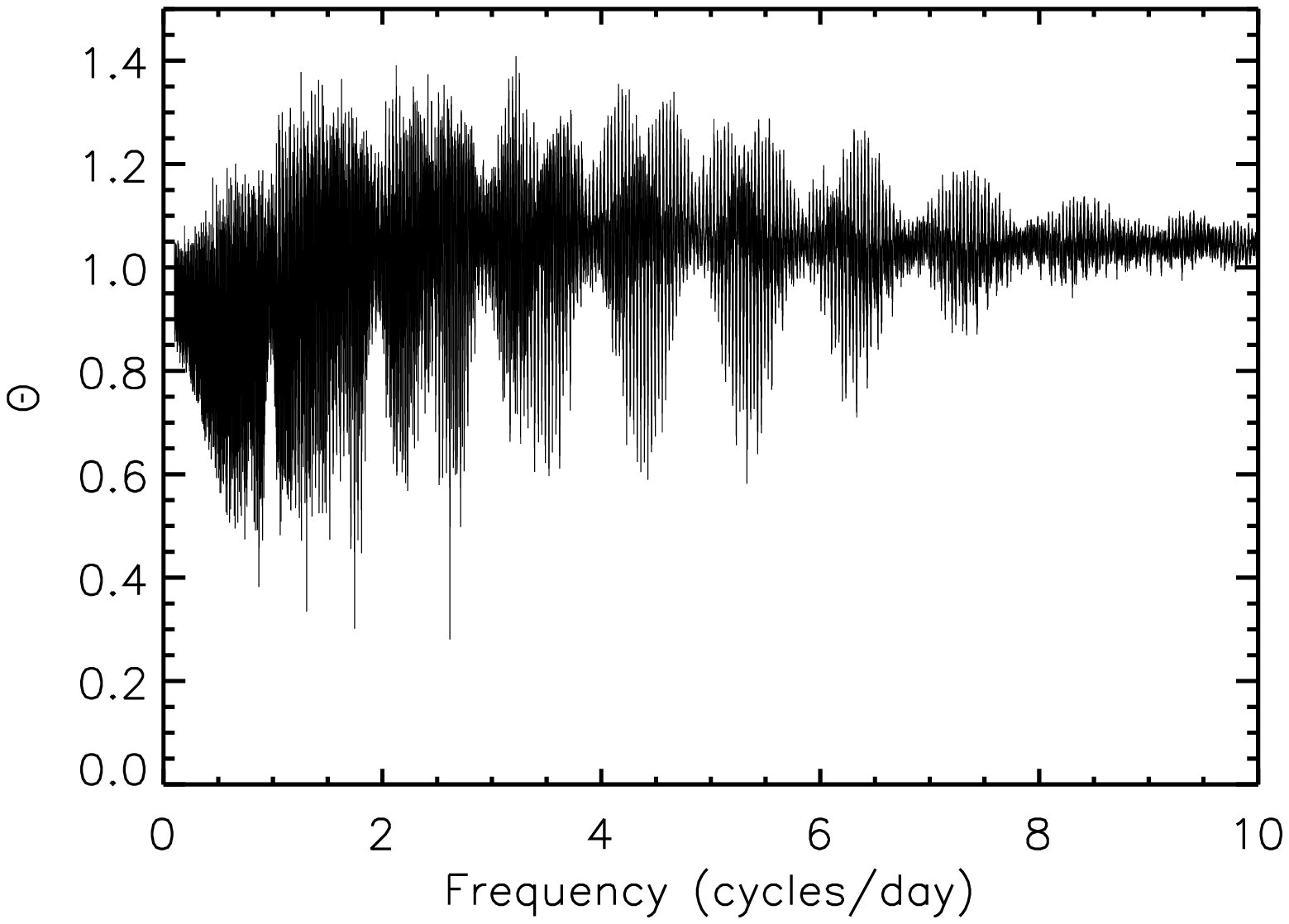}
        \caption{PDM}
        \label{fig:PDM}
    \end{subfigure}%
    \caption{(a) Lomb periodogram spectral power derived from 2014, 2015, and 2019 photometric data. The spectral power is plotted as a function of the frequency (cycles/day). The maximum spectral power of 106 is at 2.61670 cycles/day, which corresponds to a rotation period of 9.1718 h. (b) PDM plot derived from 2014, 2015, and 2019 photometric data. The minimum value of $\theta$ is given for a frequency of 2.61669 cycles/day (i.e., $P=9.17188$~h).}
    \label{fig:period_search}
\end{figure*}

Additionally, two runs were carried out with the 1.5-m telescope at Sierra Nevada Observatory, in Granada (Spain). The first run was on October 3, 24, and 25, and the second run was on November 24, and 25. This telescope has a $2{\rm k}\times2{\rm k}$ CCD with a FoV of $7.92'\times7.92'$ and an pixel scale of $0.232''/$pixel. This FoV is not large enough to aim the telescope at the same coordinates over the whole run. However, the duration of the observations each night was long enough to cover more than $50\%$ of the rotational light curve most of the nights (i.e., between 2 and 5.1 hours per night), which allows adding an offset to put all data at the same level. The average seeing of each run was $1.83''$ and $2.18''$, for the first and second run, respectively.

All runs were executed using binning $2\times2$, and without filter, in order to maximize the S/N. We took bias and flat-field frames at the beginning of each observation night. We calculated a median bias that was subtracted to the flat-field frames; after bias subtraction, a normalized-median flat-field was also calculated \citep[these procedures were carried out using the software {\it AstroImageJ};][]{Collins2017}. Then, we subtracted the median bias and divided by the normalized-median flat-field each science image using specific routines written in Interactive Data Language (IDL). These routines also contain the specific code to perform the synthetic aperture photometry.

\begin{figure}

	\includegraphics[width=\columnwidth]{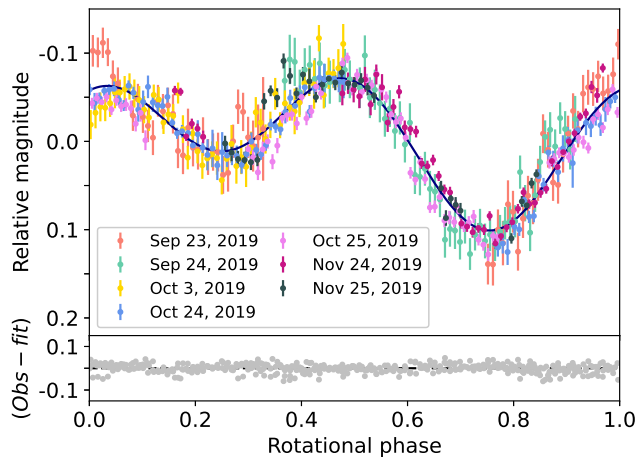}
    \caption{Bienor's rotational light curve was obtained using data from the 1.23-m and 1.5-m telescopes (see Section \ref{sec:RLC_observations} for a detailed explanation on the telescopes used here). The data were folded using a rotational period of 9.1719 $\pm$ 0.0002 h, being the zero rotational phase at the time of the occultation. Different colours represent different observation nights as described by the legend. The blue line represents a second-order Fourier-function fit to the observational data, the peak-to-valley amplitude of the fit has a value of $0.172\pm0.003$ mag. At the bottom panel, gray dots represent the residual of the fit to the observational data, with a standard deviation of 0.02 mag.}
    \label{fig:RLC_2019}
\end{figure}

We applied the same photometric techniques as in \cite{Fernandez-Valenzuela2016}. All the stars used in the photometry were brighter than Bienor so that the main source of errors comes from Bienor's flux. In order to maximize the quality of the photometry, we studied the data dispersion versus aperture radii for each night, selecting those radii that minimize the dispersion.

Before calculating the rotational phase at the time of the occultation, we refined the rotation period of Bienor by merging these data with those from previous campaigns taken in 2014 and 2015. Results from these campaigns were previously published in \cite{Fernandez-Valenzuela2017}, where the relative photometric measurements are available online. All data were corrected from light-travel time. We used two different techniques to search for the rotation period in unevenly sampled data. The first one is the Lomb technique \citep{Lomb1976}, which is a modified version of the Fourier spectra analysis as implemented in \cite{Press1992}. The resulting periodogram can be seen in Figure \ref{fig:Lomb}, where the maximum spectral power results for a rotation period of 9.1718 h. The second method is the Phase Dispersion Minimization (PDM) technique (see Figure \ref{fig:PDM}), which searches for the period that minimizes the value of the so-called $\theta$ parameter \citep{Stellingwerf1978a}. The minimum is found for a rotation period of 9.17188~h.  

We folded our 2019 data set using the resulting rotation periods from both techniques. Bienor's rotational light curve is thought to be due to a tri-axial body shape, as previously stated in the literature \citep{Ortiz2003,Rabinowitz2007,Fernandez-Valenzuela2017}; therefore, the set of data was fitted to a second-order Fourier function as follows:
\begin{equation}
\label{eq:Fourier_function}
    m=\Sigma_{i}\left[a_i\sin(2i\pi\phi)+b_i\cos(2i\pi\phi)\right],
\end{equation}
where $m$ is the theoretical value of the relative magnitude obtained from the fit, $\phi$ is the rotational phase, and $(a_i$, $b_i)$ are the coefficients of the Fourier function (with $i=0,1,2$). The rotational phase is calculated as the fractional part of $({\rm JD}-{\rm JD_0})/P$, where JD is the Julian date, JD$_0=2458494.46234585$ is the initial Julian date (which corresponds to the moment of the occultation, corrected from travel-light time), and $P$ is the rotation period (in days). We searched for the best fit among the values obtained from the different period-search routines. The goodness of the fit was evaluated using the $\chi^2_{\rm PDF}$ test, where the solution that minimizes the value of $\chi^2_{\rm PDF}$ (with $\chi^2_{\rm PDF}>1$) for the whole set of data (2014, 2015, and 2019) is $9.1719\pm0.0002$ h. The uncertainty of the rotation period is obtained by searching for the value that produces a $\chi^2_{\rm PDF}+1.61$ \citep[from the $\chi^2$ distribution with a 0.9 level of confidence for 5 degrees of freedom;][]{Snedecor1989}. The resulting rotational light curve from 2019 data can be seen in Figure \ref{fig:RLC_2019}. 

The peak-to-valley amplitude (i.e., the difference between the brightness absolute maximum and minimum), $\Delta m$, of the rotational light curve is $0.172\pm0.003$ mag. This value is larger than those from years 2014, 2015, and 2016 reported in \cite{Fernandez-Valenzuela2017}, and in agreement with the models therein that predict the increment in $\Delta m$ for future years due to the variation of the aspect angle (the angle between the rotation axis and the line of sight) of the body. We used the instant at which the stellar occultation occurred as initial Julian date. Therefore, it is straightforward to observe that the event took place very close to a maximum of brightness.

%------------------------------------------------------------------------------------------------------------------------

%------------------------------------------Data Analysis ----------------------------------------------------------------

\section{Analysis}
\label{sec:analysis}

\begin{figure*}
\centering
	\begin{subfigure}{.48\textwidth}
        \centering
        \includegraphics[width=\linewidth]{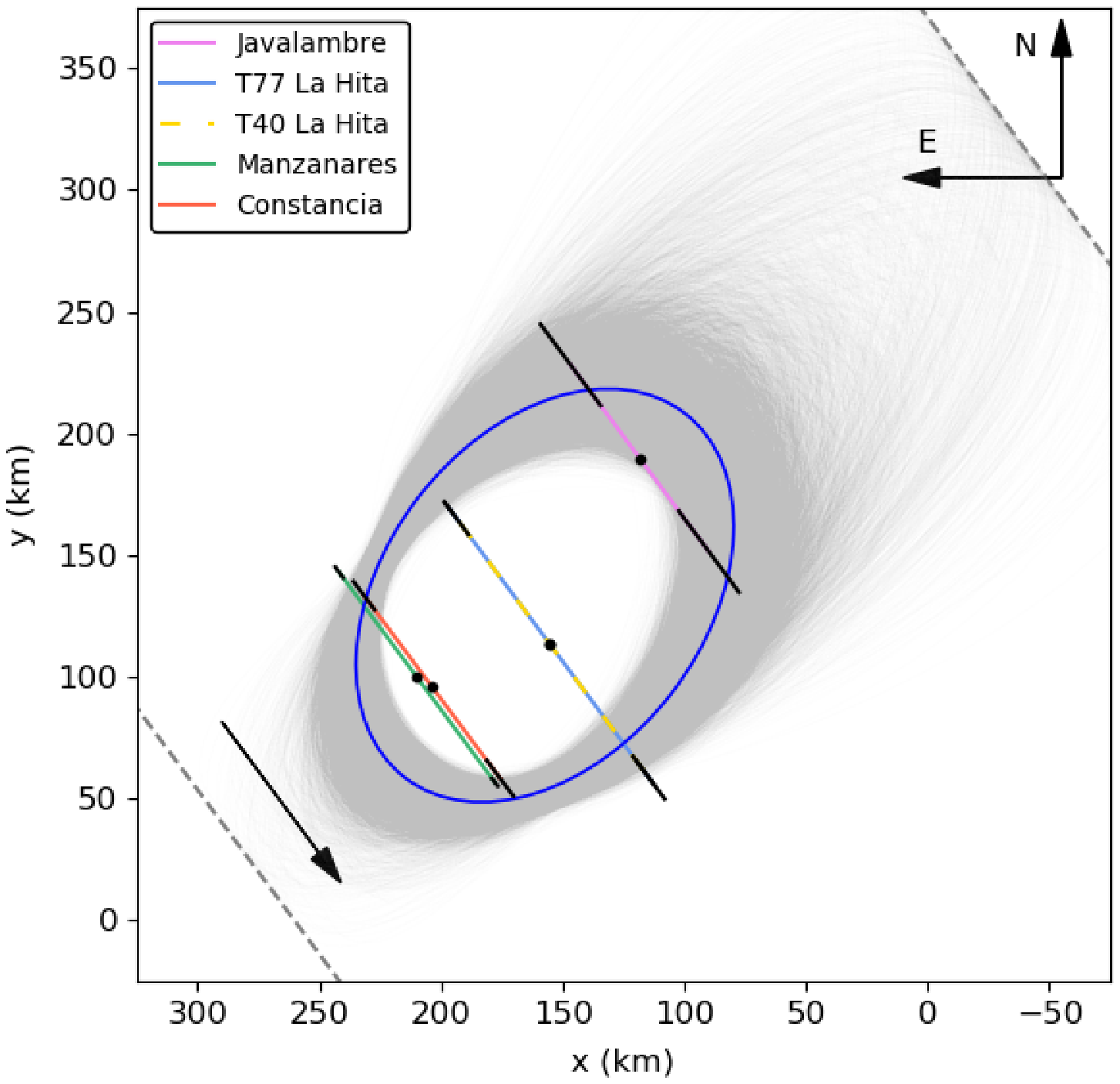}
        \caption{S1}
        \label{fig:chords_Manzanares_displaced}
    \end{subfigure}%
	    \begin{subfigure}{.48\textwidth}
        \centering
        \includegraphics[width=\linewidth]{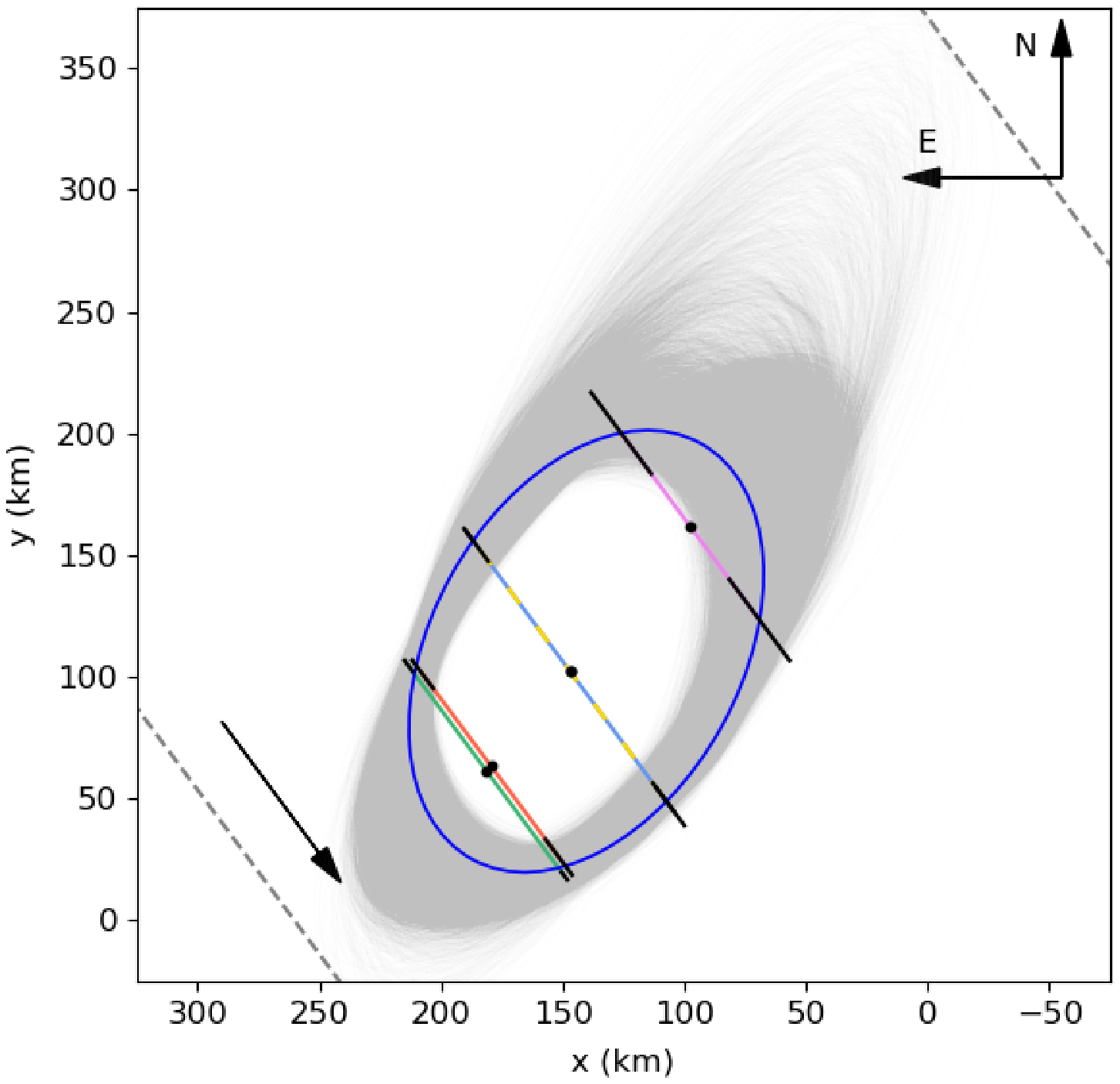}
        \caption{S2}
        \label{fig:chords_all_aligned}
    \end{subfigure}%
\caption{Elliptical fit to the five occultation chords obtained from the light curves shown in Figure \ref{fig:Occultation_LC}, where $x$ and $y$ depict the distances in the sky plane in the East-West and North-South directions respectively. The gray shadowed region represents the 10,000 ellipses given by the Monte Carlo method, with the blue ellipses representing the best fit to the extremities of the chords by least-squares minimization of the geometric distances. For reference, the colours of the chords match those of Figure \ref{fig:Occultation_LC}, with the black solid lines in the extremities representing the 1-$\sigma$ uncertainties of the ingress/egress times. We note that La Hita T40's and T77's chords are overlapping one another. gray dashed lines outside the ellipse represent the closest locations (to the North and South of the object) where the stellar occultation was not detected, which correspond to the data from Agrupaci\'on Astron\'omica de Sabadell (Sabadell) and La Sagra (Granada). Black arrows at the left-bottom corner indicate the star's motion relative to the centaur. (a) Manzanares's chord is displaced to match Constacia's ingress (S1 fit). (b) All chords' centers are aligned (S2 fit).}
    \label{fig:chords}
\end{figure*}

In order to determine Bienor's projected shape, we translated the ingress and egress times from the light curves (Figure \ref{fig:chords}) into chords, in other words, the projected lines on the sky's plane as observed from each location. The extremities of these chords constrain Bienor's projected shape at the time of the occultation. When plotting all chords together, we noticed that the easternmost chords in the figure (Constancia and Manzanares) were displaced one from each other by $\sim50$ km. This likely indicates that the absolute times (UTC) of the recording (FITS images or video) of one or both PC's are probably not correct, in other words, the PC (system) time had an (unknown) offset. While Constancia's camera used GPS-inserted time stamps overlaid on the video images, Manzanares' instrumentation involved a laptop synchronized using an NTP connected to an internet network through a smartphone. Therefore, we regard the sharp ingress time of Constancia's observation as the correct one. We note that Constancia's chord presents high data dispersion (see Figure \ref{fig:Occultation_LC}), especially at egress, which could be affecting the calculation of the ingress and egress timings. However, the light curve shows a clear start point of the occultation, with a drop in brightness from which the timing is accurately obtained. Regarding the duration of the event, both chords present similar lengths and are in agreement considering the error bars (see Table \ref{tb:ingress_egress}).

Considering all of the above, we have evaluated two different scenarios. A first scenario (S1) where Manzanares' chord is shifted in order to match its ingress with that of Constancia (Figure \ref{fig:chords_Manzanares_displaced}). Such displacement had a value of 55~km (4~s). In a second scenario (S2), we have decided to align the centers of the chords as done in other stellar occultation works, to account for possible needed time-shifts \citep[e.g.,][]{Vara-Lubiano2022, Santos-Sanz2022}. We performed a weighed linear fit to the centers of all the chords, and then shifted all the chords so that their centers would lay on that linear fit. The shifts applied to the chords for this second scenario are included in Table \ref{tb:ingress_egress}.

As stated above (see Section \ref{sec:RLC_observations}), Bienor is expected to have a triaxial shape at least to a first approximation. Therefore, we fitted the extremities of the chords to an ellipse. To evaluate the effect of the uncertainties of the ingress and egress times, we used a Monte Carlo method: we repeated the fit $10,000$ times changing the extremities of the chords according to their uncertainties and fitting the best ellipse using the {\it curve\_fit} function of Python \texttt{SciPy}, which minimizes the sum of the squares of the geometric distances $r$ between the fitted ellipse and the extremities following the direction of the chords:
\begin{equation}
   S= \sum_i r_{\rm i}^2,
\end{equation}
with $i=1,2,\dots,N$; being $N$ the total number of extremities (i.e., twice the number of chords). ``Negative'' chords (gray lines outside elliptical figures) correspond to light curves where no occultation is detected. These negative chords put limits on the size of the fitted ellipse and the uncertainty region of that ellipse (see Figure \ref{fig:chords}, dashed gray lines). 

We performed a total of 10,000 Monte Carlo iterations that produced different ellipses defined by the following parameters: semi-major and -minor axes ($a_{\rm p}$ and $b_{\rm p}$, respectively), tilt angle ($\theta$, from the abscissa's positive axis to the ellipse's semi-major axis, clockwise as drawn in Figure \ref{fig:chords}), and center's coordinates ($x$, and $y$). The probability density function (PDF) for each parameter can be seen in Appendix \ref{app:probability_density_functions}. The results are displayed in Figure \ref{fig:chords}, where the blue lines represent the best fit to the extremities of the chords and the gray shadowed regions represent all 10,000 ellipses from the Monte Carlo method. Table \ref{tab:results_from_fits} shows the resulting values for the ellipses' parameters with the errors provided by the standard deviation of the PDF given by the Monte Carlo modelization.

In both scenarios, Bienor's mean area-equivalent diameter ($D'_{\rm eq}$) results in $150\pm20$ km\footnote{We note that the instantaneous-projected areas that result from the fit have been corrected by the rotational light curve amplitude at the moment of the occultation, which is $0.058\pm0.003$ mag, in order to obtain the equivalent diameter. This value is the theoretical relative magnitude given by the Fourier fit for the rotational phase during the stellar occultation.}. This value is 30 km smaller than the one obtained from Herschel and ALMA \citep[$181\pm8$ km,][see Table \ref{tab:results_from_fits}]{Lellouch2017}. 
We obtained the albedo ($p_{\rm V}$) of the object by using the known equation:
\begin{equation}
    D'_{\rm eq} = Cp_{\rm V}^{-1/2}10^{-H_{\rm V}/5},
\end{equation}
where $C=1330\pm18$ km is a constant \citep{Masiero2021}, and $H_{\rm V}=7.47\pm0.04$ mag is Bienor's absolute magnitude (rotationally averaged) in $V$-band \citep[from][]{Fernandez-Valenzuela2017}.

\begin{table*}
    \caption{Parameters of the ellipses and physical properties obtained from the fit.}
    \label{tab:results_from_fits}
    \centering
    \begin{tabular}{lcccccccc}
    \hline \hline
    Scenario    & $a_{\rm p}$   & $b_{\rm p}$   & $\theta$  & $x$   & $y$   & $D_{\rm eq}$ & $D'_{\rm eq}$  &$p_{\rm V}$ \\ 
                & (km)          &(km)           & ($^{\circ}$) & (km)   &(km) & (km)& (km)&(\%)\\
    \hline
    & & & & & & & & \\

  S1 & $94\pm27$ & $66\pm5$ & $128\pm13$   & $157\pm15$ & $133\pm15$  & $160\pm20$         & $150\pm20$ & $8\pm2$\\\vspace{-2.5mm}
 & & & & & & & &\\
  S2  & $97\pm24$ & $64\pm4$ & $119\pm7$   & $140\pm8$ & $110\pm14$& $160\pm20$ & $150\pm20$        & $8\pm2$\\ 
  & & & & & & & & \\
   \hline
     & & & & & & & & \\
  S3 E1 & $84\pm6$ & $57\pm4$  & $8\pm10$  & $158\pm8$ & $112\pm3$ & \multirow{2}{*}{$188\pm9$\tablefootmark{\dag}} & \multirow{2}{*}{$183\pm8$\tablefootmark{\dag}} & \multirow{2}{*}{$5.7\pm0.6$\tablefootmark{\ddag}}\\\vspace{-2.5mm}
    & & & & & & & & \\
S3 E2 & $67\pm8$& $60\pm6$ & $7\pm10$ & $83\pm10$ & $210\pm9$ & & & \\
  & & & & & & & & \\
        \hline \hline
    \end{tabular}
    \tablefoot{{Abbreviations are defined as follows: fit after displacement of Manzanares chord (S1), fit aligning centers of all chords (S2), fit to a binary system (S3, with E1 and E2 indicating the easternmost and westernmost ellipse, respectively), semi-major and minor axes of the projected ellipse on the sky ($a_{\rm p}$ and $b_{\rm p}$, respectively), tilt angle ($\theta$), center of the ellipse ($x$,$y$), instantaneous area-equivalent diameter ($D_{\rm eq}$), rotationally averaged area-equivalent diameter ($D'_{\rm eq}$), and albedo ($p_{\rm V}$). We note that other ellipse fitting methods give similar ellipse parameters with considerably smaller uncertainties than the Monte Carlo method, and thus the difference in equivalent diameter of Bienor with respect to the thermal models becomes even more significant. However, we prefer to show the more conservative results here. 
}
\tablefoottext{\dag}{Total area-equivalent diameter considering both bodies.}\tablefoottext{\ddag}{Albedo considering the total area-equivalent diameter of both bodies.}
    }
\end{table*}

However, the position angle ($PA$\footnote{The angle formed by the North and the projected pole orientation (i.e., rotational axis) on the sky, toward East.}) of the short axis of the ellipse resulting from these fits is not in agreement with the one calculated using Bienor's pole orientation from \cite{Fernandez-Valenzuela2017}. Since the stellar occultation happened very close to a maximum of brightness of Bienor's rotational modulation (see Figure \ref{fig:RLC_2019}), we can assume that the semi-minor axis of the ellipse ($b_{\rm p}$) is a proxy of the projection of the pole orientation of the body (as have been the cases of Chiron, Chariklo, and Haumea); therefore, $PA=180^{\circ}-\theta$. For the different scenarios presented here, $PA$ is $52^{\circ}$, and $61^{\circ}$, for S1, and S2, respectively (or, their supplementary direction, $232^{\circ}$, and $241^{\circ}$). Bienor's pole orientation was constrained in \cite{Fernandez-Valenzuela2017} using long-term photometric models, which provided a preferable range of values of the ecliptic longitude ($\lambda_{\rm p}$) and latitude ($\beta_{\rm p}$) of $[25^{\circ},40^{\circ}]$ and $[45^{\circ},55^{\circ}]$, respectively, under the assumption of hydrostatic equilibrium. If we include models out of the hydrostatic equilibrium, $\beta_{\rm p}$ reaches up to $60^{\circ}$. Although it is a small difference of only $5^\circ$, we included it in our calculations. The pole orientation can be translated to the object's $PA$ at the moment of the stellar occultation by means of the equation:
\begin{equation}
    PA = \arctan\left(\left|\frac{\sin(\Phi)\cos(D_{\rm p})}{\cos(D)\sin(D_{\rm p})-\sin(D)cos(D_{\rm p})\cos(\Phi)}\right|\right),
\end{equation}
where $\Phi=RA_{\rm p}-RA$, with $RA$ and $D$ being the right ascension and declination of Bienor-centered reference frame, respectively, and $RA_{\rm p}$ and $D_{\rm p}$ being the right ascension and declination of Bienor's pole orientation, respectively. Translating the above mentioned range of the ecliptic longitude and latitude of Bienor’s pole orientation to right ascension and declination, we obtained that $PA$ should be in the range $[300^{\circ},340^{\circ}]$, which is not in agreement with the short axis from the stellar occultation ellipse fits. 

\begin{figure}
        \centering
        \includegraphics[width=\linewidth]{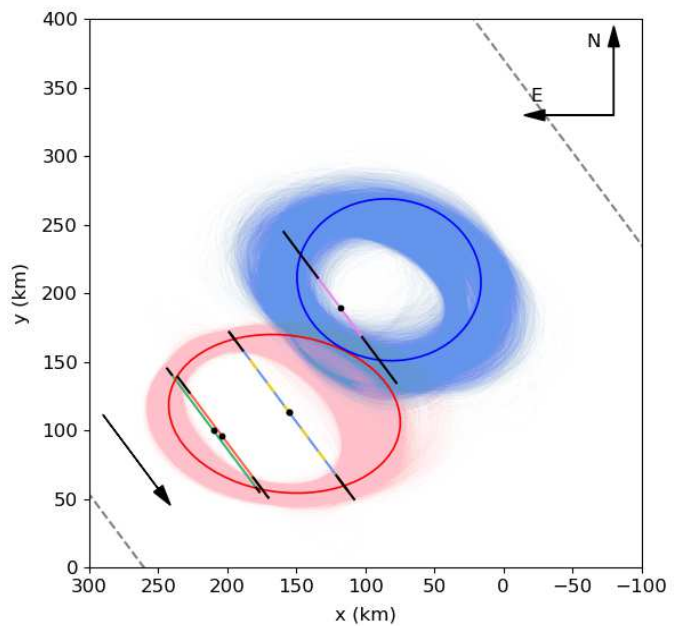}
    \caption{Simultaneous fit of two ellipses to the five chords from the light curves in Figure \ref{fig:Occultation_LC} where $x$ and $y$ depict the difference between the JPL orbit and the center of the shadow. (No shift to the chords has been applied other than Manzanares' due to the error in the computer synchronization). The blue and red ellipses represent the best fit in terms of minimization of a $\chi^2$ function, with the blue and red shadowed regions representing all 10,000 possible solutions given by the Monte Carlo method. For reference, the colours of the chords match those of Figure \ref{fig:Occultation_LC}, with the black solid lines in the extremities representing the 1-$\sigma$ uncertainties of the ingress/egress times. We note that La Hita T40's and T77's chords are overlapping one another. gray dashed lines outside the ellipses represent the closest locations (to the North and South of the object) where the stellar occultation was not detected, which correspond to the data from Agrupaci\'on Astron'omica de Sabadell (Sabadell) and La Sagra (Granada). The black arrow at the left-bottom corner depicts the shadow motion.
    \label{fig:binary}}
\end{figure}

Considering the differences found between our effective diameter and the thermal one, and the different pole orientations, we propose an hypothetical scenario (S3), in which Bienor is a close-in binary system as shown in Figure \ref{fig:binary} (where no shift to the chords has been applied other than Manzanares' due to the error in the computer synchronization). We note that this fit of two ellipses with only four chords is a degenerated problem and there exists many solutions. Therefore, in order to obtain a unique solution we need to include additional constraints. Hence, we constrained the area-equivalent diameter of the combination of the two bodies to be within the range provide by \cite{Lellouch2017}, that is $181\pm8$ km. Additionally, the position angle of the two bodies must be similar because large differences would result in a chaotic system, in other words, a tumbling system (objects with no fixed axis of rotation), which is only seeing in very small objects and would produce irregular rotational light curves in term of shape and rotation period, specially at different aspect angles. For this we chose the pole orientation to differ no more than $20^{\circ}$. This configuration results in two bodies with a total equivalent diameter of $183\pm8$ km, and an albedo of $5.7\pm0.6\%$. The resulting tilt angle for each object is $8^{\circ}\pm10^{\circ}$ and $7^{\circ}\pm10^{\circ}$, which translates into $PA$ of $172^{\circ}\pm10^{\circ}$, and $173^{\circ}\pm10^{\circ}$ (with supplementary directions of $352^{\circ}\pm10^{\circ}$ and $353^{\circ}\pm10^{\circ}$) for the easternmost and westernmost object, respectively. Therefore, this scenario allows for a value of $PA$ in agreement with the above interval obtained from the pole orientation given in \cite{Fernandez-Valenzuela2017}.

\section{Discussion}
\label{sec:Discussion}

Given Bienor's size, the possibility of Bienor being out of the hydrostatic equilibrium is reasonable, and its shape could depart considerably from a tri-axial ellipsoid. This possibility is considered in the first scenario, S1, where the chords display a somewhat irregular shape for the body. However, with chords at only three locations of the body we are limited to performing an elliptical fit in order to obtain some physical properties. Another possibility could be that Bienor is in fact in hydrostatic equilibrium and therefore we would expect for all chords to have their center aligned within error bars. This would imply that the displacements among the different chords are due to various systematic errors (as those associated with the computer synchronization). This possibility is evaluated in the second scenario (S2) where we have aligned the centers of all chords.

\begin{figure}
	\includegraphics[width=\columnwidth]{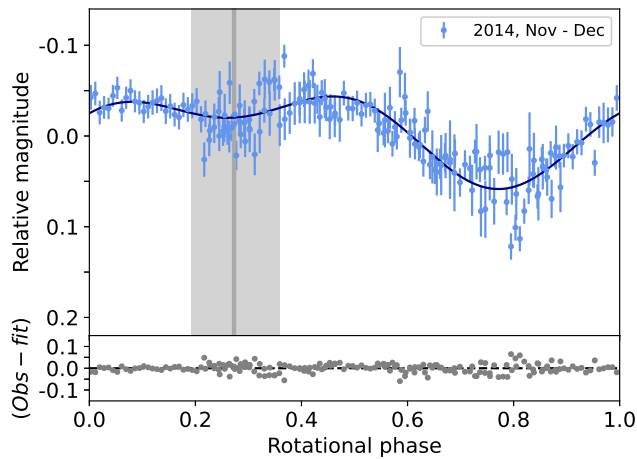}
    \caption{Bienor's rotational light curve taken in 2014. The data was folded using a rotation period of $9.1719\pm0.0002$ h, being the zero rotational phase at the moment of the stellar occultation. Light blue dots represent observation data and the dark blue line represents a second-order Fourier-function fit, with an amplitude of $0.088\pm0.007$ mag. The gray vertical line indicates the rotational phase of the two Herschel measurements, with the gray shadowed region representing the error. At the bottom panel, gray dots represent the residual of the fit to the observational data. \citep[Adapted from][]{Fernandez-Valenzuela2017}.
    }
    \label{fig:Herschel_rotation_phase}
\end{figure}

However, as mentioned in Section \ref{sec:analysis}, both scenarios produce an area-equivalent diameter 30 km smaller than that obtained from thermal models, which is $181\pm8$ km according to \cite{Lellouch2017}, and $198^{+6}_{-7}$ km according to \cite{Duffard2014}. In order to know if this difference is due to Bienor's rotational modulation, we calculated Bienor's rotational phase at which Herschel measurements were obtained\footnote{Mid-time of exposure of Herschel images: 2011-01-24T12:52:20 and 2011-01-24T22:00:27.}, using the highly accurate rotation period obtained in Section \ref{sec:RLC_observations}, and the closest published rotational light curve in time to the Herschel measurements, which was taken in 2014 \citep{Fernandez-Valenzuela2017}. Two flux measurements were taken with Herschel and both at the same rotational phase, during the brightest minimum (gray vertical line in Figure \ref{fig:Herschel_rotation_phase}), with an estimated uncertainty of 0.768~h (i.e., $\sim0.08$ in rotational phase, gray shadowed region in Figure \ref{fig:Herschel_rotation_phase}). Since Herschel measurements were taken at a minimum of brightness, and due to the small amplitude of the rotational light curve, this size difference cannot be explained by rotational variability. Neither can be due to a change in the aspect angle because the amplitude of the rotational light curve has been increasing in recent years ($0.083\pm0.008$ mag from 2014 to 2019), which implies that the object has been changing from an pole-on configuration to a more edge-on configuration.

On the other hand, the position angles provided by the fits for S1 and S2 might be incompatible with the results from the long-term photometric models in \cite{Fernandez-Valenzuela2017}. In other words, the pole orientation obtained using the rotational light curves taken at different aspect angles could be incompatible with the results of the fit from the stellar occultation data when using a single object. This is an unexpected outcome because results from this kind of models for other objects match those obtained through stellar occultations. For instance, the stellar occultation by Chariklo detected in 2013 produced two sets of values for the pole orientation: $(\lambda_{\rm p},\beta_{\rm p})=(138^{\circ},28^{\circ})$ or $(\lambda_{\rm p},\beta_{\rm p})=(26^{\circ},-7^{\circ})$, see \cite{Braga-Ribas2014}. Later, \cite{Duffard2014b} discarded one of the solutions using long-term photometric models \citep[see figure 1 in][]{Duffard2014b}. A similar situation appears in the case of Chiron, where two pole orientations were obtained through a set of stellar occultations and one of them fits perfectly the long-term photometric data \citep{Ortiz2015}.

A close-in binary system is a hypothetical scenario in which we can match the results from the stellar occultation with the long-term from photometric models published in \cite{Fernandez-Valenzuela2017} and that also allows for a larger equivalent diameter, more compatible with the radiometric derived one. We can speculate that the system could be interacting by tidal forces shrinking the orbit until it forms a contact binary, similar to the formation scenario proposed for Arrokoth \citep{McKinnon2020}, the TNO contact binary visited by New Horizons spacecraft in January 2019. Additionally, this scenario results into an area-equivalent diameter of $166\pm9$ km, more compatible with that obtained from thermal measurements \citep[$181\pm8$ km;][]{Lellouch2017}.

A synchronized binary system would also explain Bienor's rotational light curve that presents an invariable shape over the years, with the variation in amplitude coming from the change in the aspect angle of the system. This could be another hint to discard an irregular object, from which one would expect different shapes of its rotational light curves when observed at different aspect angles. The synchronization of spin and orbit of both components is a reasonable assumption since the time required to achieve this status is much smaller than the age of the Solar System ($1.5\times10^{17}$ s), as we demonstrate in the following. The time required by one of the components to be tidally lock, $t_{\rm lock}$, can be obtained using the following equation \citep{Hubbard1984}:
\begin{equation}
t_{\rm lock} = \frac{2\pi}{3K\,G\,a_{\rm 1}^3T_0 \delta}\left(\frac{d^6M_{1}}{M_{2}^2}\right).
\end{equation}
In this equation, $K$ is the secular Love number, which has a value of 3/2 for homogeneous bodies, $G$ is the gravitational constant, $a_{\rm 1}$ and $T_0$ are the semimajor axis and the initial rotation rate of the largest component, respectively, $\delta$ is expressed as $\arctan(1/Q)$ with $Q=100$ being the dissipation factor, $d$ is the distance at which the smaller component orbits the largest component, and $M_{1}$ and $M_{2}$ are the masses of each component.

The distance at which the secondary orbits the primary, can be obtained from the equation:
\begin{equation}
d^3=\frac{T_{\rm s}^2G\,M_{\rm T}}{4\pi},
\end{equation}
where $T_{\rm s}$ is the orbital period of the secondary, and $M_{\rm T}$ is the mass of the system. Combining both equations and assuming that both components have similar masses and radii (i.e., $M_1\approx  M_2 = M$, and $R_1 \approx R_2=R$, respectively), the time for the system to be tidally locked can be obtained from:
\begin{equation}
t_{\rm lock} = \frac{T_{\rm s}^4G\,M}{6\pi\,T_0K\,a_1^3\delta},
\end{equation}
with $M=V\rho= 4/3\pi R^3\rho$, where $R$ is the effective radius and $\rho$ is the density, which considering centaurs of similar sizes, is 700 kg/m$^{-3}$. Therefore, considering $T_0\geq9.1718$ h (Bienor's rotation period at the present time):
\begin{equation}
t_{\rm lock} \lesssim 2.5\times 10^7\,{\rm G s},
\end{equation}
which is orders of magnitude smaller than the age of the Solar System.

We note that the Roche limit, given by the equation:
\begin{equation}
d_{\rm R} = 2.44a_1\left(\frac{\rho_1}{\rho_2}\right)^{1/3}\approx2.44 a_1,
\end{equation}
with $a_1=84$ km (from S3 E1, see Table \ref{tab:results_from_fits}), results in 205 km, which is somewhat larger than the distance between the centers of the two bodies measured in Figure \ref{fig:binary}, $d=150\pm20$ km. However, as the binary fit is degenerated other possible solutions exist with larger separation between the bodies. On the other hand, if the body has some internal strength the components can be closer-in than the nominal Roche limit for fluid-like bodies.

\section{Limits for a putative ring system}

\label{sec:ring-satellite_discussion}

The existence of a ring system would also provide a better match between the area-equivalent diameter from the stellar occultation versus thermal models. This scenario was proposed by \cite{Fernandez-Valenzuela2017} because of the resemblance of Bienor in size, shape, and watery spectra to Chariklo and Chiron \citep[for which, although still under debate, it seems very plausible to possess a ring system, as can be seen in][]{Ortiz2015,Sickafoose2020}. The long-term photometric models presented therein worked considerably well, and in fact, allowed for smaller sizes in the main body (in agreement with the area equivalent diameters derived in this work), since the potential ring would be contributing to a large portion of the brightness, even more so if the water-ice detected in Bienor's spectra \cite[e.g.,][]{Dotto2003,Barkume2008} is located in the ring, as in the case of Chariklo \citep{Duffard2014b}. It has been claimed by several authors that centaurs present a colour bimodality that split the population between very red objects and less red objects \citep{Thebault2003,Peixinho2012}, with median albedos of $\sim8\%$ and $\sim5\%$, respectively \citep{Mueller2020}. Although studies using absolute magnitudes, which are not affected by the phase angle at which the observations are taken, do not reflect such a bimodality \citep{Alvarez-Candal2019}. In any case, the correlation between very redness and albedo is still present on centaurs. With a $V-R$ colour of 0.44 mag \cite[see, e.g.,][]{Delsanti2001,Doressoundiram2002}, Bienor is one of the less red centaurs, which would suggest a somewhat lower albedo for Bienor, contrary to what we obtain (see Table \ref{tab:results_from_fits}). A watery ring around Bienor would decrease the albedo of the main body to smaller values and would explain the abrupt increase in brighness from 2000 to 2013 noticed in \cite{Fernandez-Valenzuela2017}. The thermal data from Herschel would be then the combination of the main body and ring reemission, resulting in larger values for the main body, which would explain the difference between the area-equivalent diameter from the stellar occultation and thermal data.

\begin{figure*}
\centering
	\begin{subfigure}{.5\textwidth}
        \centering
        \includegraphics[width=\linewidth]{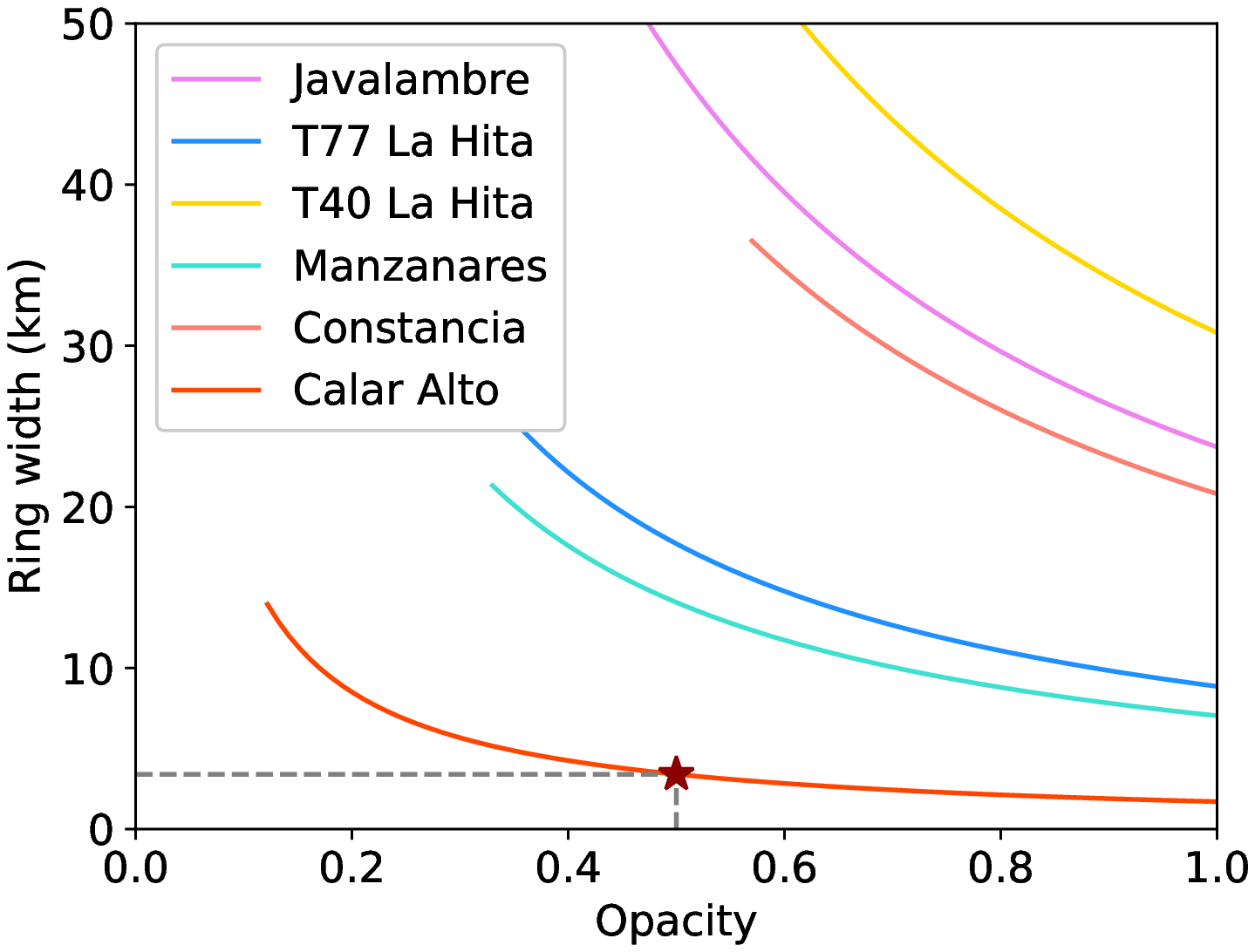}
        \caption{}
        \label{fig:ring_constraints}
    \end{subfigure}%
	    \begin{subfigure}{.45\textwidth}
        \centering
        \includegraphics[width=\linewidth]{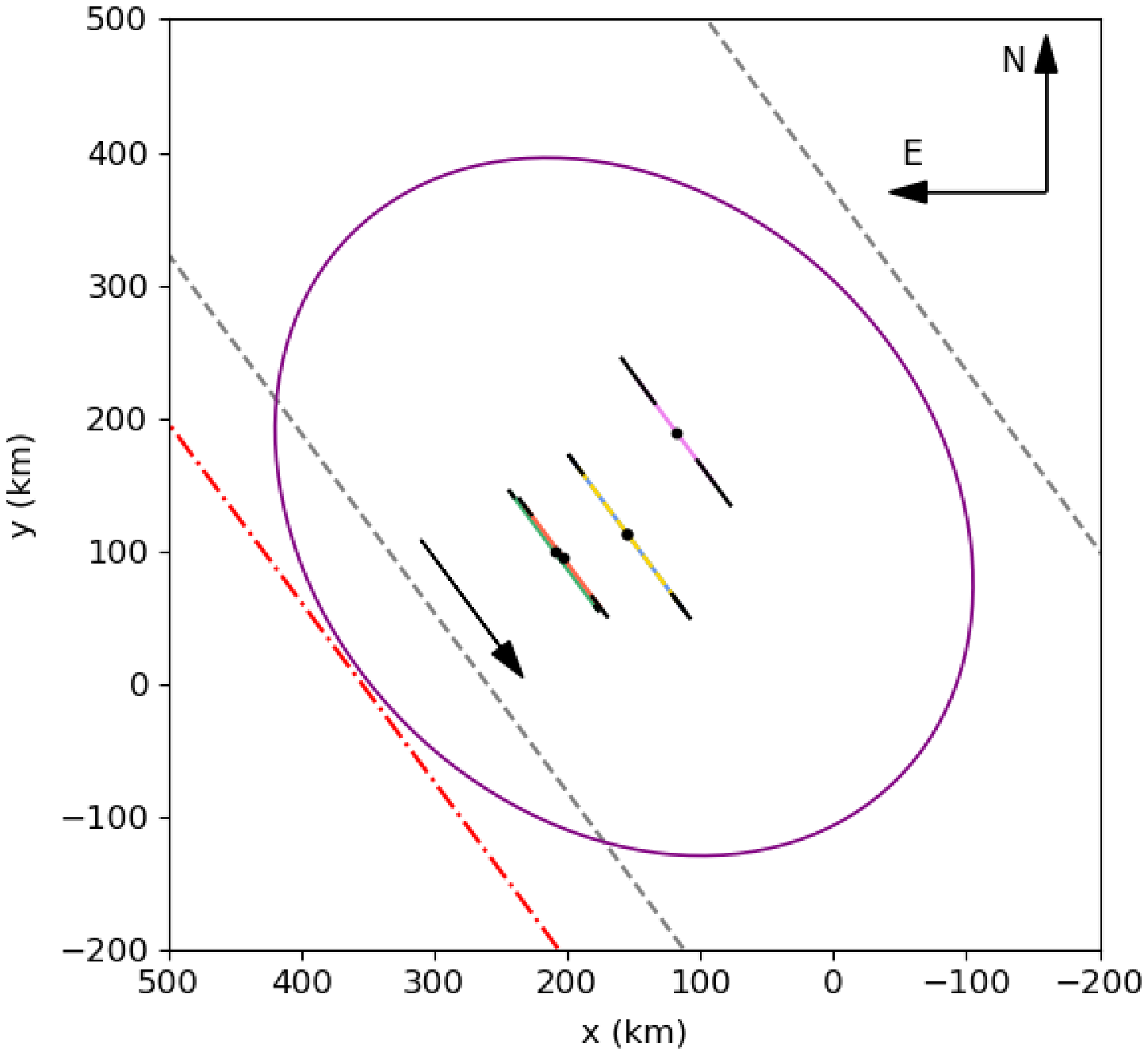}
        \caption{}
        \label{fig:possible_ring}
    \end{subfigure}
    \label{fig:rings}
    \caption{(a) Lower limit for the ring width that would be detected considering the dispersion of each light curve and the corresponding instrumentation used in this work. Our best constraint is given by Calar Alto's data (red line). The brown star indicates the lower limit (3.4 km) for a ring that would be detected with an opacity of 50\%. (b) Positive and negative chords (as plotted in Figure \ref{fig:chords}a) including the negative chord from Calar Alto (red pointed-dashed line) that would have detected the ring system. A similar ring in width to those found in Chariklo and Chiron is plotted in purple with the expected position angle and aspect angle determined from the body pole axis orientation in \cite{Fernandez-Valenzuela2017}. A ring with larger diameter would be outside the Roche limit for plausible densities of Bienor. The black arrow indicates the motion of the shadow.}
\end{figure*}

Assuming Bienor is currently near to a pole-on orientation as the most plausible explanation for the decrease in the amplitude of the rotational light curve during the last 20 years, the area of the putative rings would be highly exposed and could have been detected in our data. Therefore, we searched within Bienor's stellar occulation data, but no features similar to those found in Chariklo and Chiron were found. However, the lack of secondary drops does not prove the lack of rings nor satellites. On the one hand, the probability of detecting a satellite around Bienor with only a handful of chords is negligible. On the other hand, if a putative ring is very narrow and/or presents lower opacity than what can be detected with the precision given by the used instrumentation, the ring could be overlooked. The lower limit of ring width ($w_{\rm l}$) under $3\sigma$ uncertainty at which a ring system is detected is given by the following equation:
\begin{equation}
    w_{\rm lim} = \frac{3\sigma\,v\,T_{\rm exp}}{\tau},
\end{equation}
where $\sigma$ is the dispersion of the light curve, $v$ is Bienor sky motion at the moment of the occultation (in km$\,{\rm s}^{-1}$), $T_{\rm exp}$ is the exposure time (in s), and $\tau$ is the opacity. Figure \ref{fig:ring_constraints} shows the dependency of $w_{\rm lim}$ with the opacity for each of the light curves in Figure \ref{fig:Occultation_LC}. We note that the minimum optical depth of a detectable putative ring is restricted by the dispersion of each light curve. We included also light curves with a negative detection of the main body since a ring's shadow should cover a larger area than that of the body. As can be seen, our strongest constraint is given by Calar Alto's data, for which a ring with a width as narrow as 1.7 km and 100\% opacity would be detected. A 100\% opacity means a very optically thick ring, which is very unrealistic. For a ring with 50\% opacity (as Chariklo's rings), rings as narrow as 3.4 km would be detected from Calar Alto. A similar ring's area to that resulting from a 3.4 km at Calar Alto distances was studied in \cite{Fernandez-Valenzuela2017}, where the optical contribution of a body plus ring can be compared to that of a single body in figure 3 of that work. However, Calar Alto location was too far from the body and a ring system similar to those of Chariklo and Chiron would be right at the limit and might have been missed (see Figure \ref{fig:possible_ring}). Considering the remaining light curves, our strongest constraint is given by Manzanares' data, but in this case, for an opacity of 50\%, a ring with $w<14.1$ km would not be detected. Therefore, from our data, we cannot discard the possibility of Bienor having a ring system or a satellite that might be influencing the photometric and thermal measurements.

Moreover, Bienor's rotational light curve might also be consistent with the existence of a ring system. As we mentioned before, Bienor presents two different depths for each of the minima (Figure \ref{fig:RLC_2019}), which is found in all other rotational light curves in the literature from 2001 to 2016 \citep[see][and Section \ref{sec:RLC_observations} in this work]{Ortiz2003,Rabinowitz2007,Fernandez-Valenzuela2017}. All of them present a difference of $\sim$0.1 mag between both minima, in other words, the difference in brightness between both minima is the same regardless of Bienor's aspect angle. For an irregular body, one might expect that a change in the aspect angle will imply a brightness variation of one minimum respect to the other, which is not the case here. A possible explanation could be an albedo spot in that part of Bienor's body, but this spot would need to cover a high range of latitudes of the body so that it appears at all aspect angles covered by the rotational light curves from 2002 to 2019. A partial ring (i.e., an arc) in a 1:1 resonance could also increase the brightness in one of the minima, with no variability at long timescales (i.e., due to the aspect angle). Dynamical simulations show that, in principle, ring-arc material might exist at 1:1 resonance in a nonsymmetrical body \citep{Sicardy2019,Sicardy2020}, with the spreading time scales increasing with the elongation of the body. Although the same simulations also demonstrate that even the dynamically stable points are unstable against dissipative collisions. In any case, the existence of the ring material cannot be confirmed from the data in this work. More stellar occultations may shed light into this matter.

%--------------------------------Conclusion-----------------------------------------------------------------------------

\section{Conclusions}
\label{sec:conclusions}

We have observed a stellar occultation by Bienor on January 11$^{th}$, 2019. We obtained five positive and four negative detections that helped to constrain the size of the body. For a single tri-axial object, we obtained area-equivalent diameters of $150\pm20$ km, smaller than that obtained from thermal measurements, and with a higher albedo of $8\pm2$\%. In scenarios S1 and S2, the resulting position angle does not match the values calculated using the pole orientation obtained from long-term photometric models \citep{Fernandez-Valenzuela2017}. To approximate the occultation results to the thermal results, a third scenario S3 has been considered, in which a close-in binary also allows for a better match with the effective diameter of $181\pm8$ km derived from thermal data. No features related to rings or satellites have been found within the data. However, a ring system similar to that of Chariklo cannot be discarded due to insufficient S/N of the observations. 

It is clear that there exist some differences among Bienor's physical properties obtained through different methods. In order to unveil the three-dimensional shape of Bienor or if it is a close-in binary system, new data need to be acquired. A new multichord stellar occultation could shed light in this regard, either revealing one of the proposed scenarios here or any other reality that we have not come up with. 

\begin{acknowledgements}
The work leading to these results has received funding from the European Research Council under the European Community's H2020 2014-2021 ERC Grant Agreement no.669416 ``Lucky Star''. E.~F.-V.~acknowledges financial support by the Space Research Initiative from State of Florida. P.~S.-S.~acknowledges financial support by the Spanish grant AYA-RTI2018-098657-J-I00 ``LEO-SBNAF'' (MCIU/AEI/FEDER, UE). A.~P.~acknowledges financial support of the Hungarian National Research, Development and Innovation Office (NKFIH) Grant K-138962. G.~B.-.R~acknowledges CAPES-PRINT/UNESP Process 88887.310463/2018-00, Project 88887.571156/2020-00. M.~A.~acknowledges financial support from CNPq grants with numbers 427700/2018-3, 310683/2017-3, 473002/2013-2, and FAPERJ grant n$^\circ$ E-26/111.488/2013. F.~B.-R.~acknowledges financial support from CNPq grant n$^\circ$ 314772/2020-0. J.I.B.~C.~acknowledges financial support from CNPq grants with numbers 308150/2016-3 and 305917/2019-6. R.~V.-M.~acknowledges financial support from CNPq grants with numbers 304544/2017-5, and 401903/2016-8. B.~M.~acknowledges financial support from CNPq grant n$^\circ$ 150612/2020-6. A.~R.-G.-J.~acknowledges financial support from FAPESP grant n$^\circ$ 2018/11239-8. J.~M.~T.-R.~research was supported by the research Grant No.~PGC2018-097374-B-I00, which is funded by FEDER/Ministerio de Ciencia e Innovaci\'on, Agencia Estatal de Investigaci\'on. This study was partly financed by the Coordena\c{c}\~ao de Aperfei\c{c}oamento de Pessoal de N\'ivel Superior - Brasil (CAPES) - Finance Code 001 and the National Institute of Science and Technology of the e-Universe project (INCT do e-Universo, CNPq grant 465376/2014-2). We acknowledge financial support by the Spanish grant AYA-2017-84637-R, and the State Agency for Research of the Spanish MCIU through the ``Center of Excellence Severo Ochoa" award to the Instituto de Astrof\'isica de Andaluc\'ia (SEV-2017-0709)''. This research has been partially funded by the Junta de Andaluc\'ia PY20\_01309 and Agencia Estatal de Investigaci\'on PID2020-112789GB-I00 projects. This research is partially based on observations collected at the Centro Astron\'omico Hispano-Alem\'an (CAHA) at Calar Alto, operated jointly by Junta de Andaluc\'ia and Consejo Superior de Investigaciones Cient\'ificas (IAA-CSIC). This research is partially based on observation carried out at the Observatorio de Sierra Nevada (OSN) operated by Instituto de Astrof\'isica de Andaluc\'ia (CSIC) and the Excalibur telescope at the Observatorio Astrof\'isico de Javalambre in Teruel, a Spanish Infraestructura Cientifico-T\'ecnica Singular (ICTS) owned, managed and operated by the Centro de Estudios de F\'isica del Cosmos de Aragón (CEFCA). Excalibur is funded with the Fondos de Inversiones de Teruel (FITE). This worked was partially carried out with observations from the Joan Or\'o Telescope (TJO) of the Montsec Observatory (OdM), which is owned by the Catalan Government and operated by the Institute for Space Studies of Catalonia (IEEC). We acknowledge Mike Kretlow for his useful comments that helped improving this manuscript.

\end{acknowledgements}

\bibliographystyle{aa} % style aa.bst
\bibliography{bibliography} % your references Yourfile.bib

\begin{thebibliography}{54}
\expandafter\ifx\csname natexlab\endcsname\relax\def\natexlab#1{#1}\fi

\bibitem[{{Alvarez-Candal} {et~al.}(2019){Alvarez-Candal}, {Ayala-Loera},
  {Gil-Hutton}, {Ortiz}, {Santos-Sanz}, \& {Duffard}}]{Alvarez-Candal2019}
{Alvarez-Candal}, A., {Ayala-Loera}, C., {Gil-Hutton}, R., {et~al.} 2019,
  \mnras, 488, 3035

\bibitem[{{Araujo} {et~al.}(2016){Araujo}, {Sfair}, \& {Winter}}]{Araujo2016}
{Araujo}, R.~A.~N., {Sfair}, R., \& {Winter}, O.~C. 2016, \apj, 824, 80

\bibitem[{{Barkume} {et~al.}(2008){Barkume}, {Brown}, \&
  {Schaller}}]{Barkume2008}
{Barkume}, K.~M., {Brown}, M.~E., \& {Schaller}, E.~L. 2008, The Astronomical
  Journal, 135, 55

\bibitem[{{Barucci} {et~al.}(2011){Barucci}, {Alvarez-Candal}, {Merlin},
  {Belskaya}, {de Bergh}, {Perna}, {DeMeo}, \& {Fornasier}}]{Barucci2011a}
{Barucci}, M.~A., {Alvarez-Candal}, A., {Merlin}, F., {et~al.} 2011, Icarus,
  214, 297

\bibitem[{{Braga-Ribas} {et~al.}(2014){Braga-Ribas}, {Sicardy}, {Ortiz},
  {Snodgrass}, {Roques}, {Vieira-Martins}, {Camargo}, {Assafin}, {Duffard},
  {Jehin}, {Pollock}, {Leiva}, {Emilio}, {Machado}, {Colazo}, {Lellouch},
  {Skottfelt}, {Gillon}, {Ligier}, {Maquet}, {Benedetti-Rossi}, {Gomes},
  {Kervella}, {Monteiro}, {Sfair}, {El Moutamid}, {Tancredi}, {Spagnotto},
  {Maury}, {Morales}, {Gil-Hutton}, {Roland}, {Ceretta}, {Gu}, {Wang},
  {Harps{\o}e}, {Rabus}, {Manfroid}, {Opitom}, {Vanzi}, {Mehret}, {Lorenzini},
  {Schneiter}, {Melia}, {Lecacheux}, {Colas}, {Vachier}, {Widemann},
  {Almenares}, {Sandness}, {Char}, {Perez}, {Lemos}, {Martinez},
  {J{\o}rgensen}, {Dominik}, {Roig}, {Reichart}, {Lacluyze}, {Haislip},
  {Ivarsen}, {Moore}, {Frank}, \& {Lambas}}]{Braga-Ribas2014}
{Braga-Ribas}, F., {Sicardy}, B., {Ortiz}, J.~L., {et~al.} 2014, Nature, 508,
  72

\bibitem[{{Brown} {et~al.}(2007){Brown}, {Barkume}, {Ragozzine}, \&
  {Schaller}}]{Brown2007}
{Brown}, M.~E., {Barkume}, K.~M., {Ragozzine}, D., \& {Schaller}, E.~L. 2007,
  \nat, 446, 294

\bibitem[{{Cikota} {et~al.}(2018){Cikota}, {Fern{\'a}ndez-Valenzuela}, {Ortiz},
  {Morales}, {Duffard}, {Aceituno}, {Cikota}, \& {Santos-Sanz}}]{Cikota2018}
{Cikota}, S., {Fern{\'a}ndez-Valenzuela}, E., {Ortiz}, J.~L., {et~al.} 2018,
  \mnras, 475, 2512

\bibitem[{{Collins} {et~al.}(2017){Collins}, {Kielkopf}, {Stassun}, \&
  {Hessman}}]{Collins2017}
{Collins}, K.~A., {Kielkopf}, J.~F., {Stassun}, K.~G., \& {Hessman}, F.~V.
  2017, \aj, 153, 77

\bibitem[{{Delsanti} {et~al.}(2001){Delsanti}, {Boehnhardt}, {Barrera},
  {Meech}, {Sekiguchi}, \& {Hainaut}}]{Delsanti2001}
{Delsanti}, A.~C., {Boehnhardt}, H., {Barrera}, L., {et~al.} 2001, Astronomy
  and Astrophysics, 380, 347

\bibitem[{{Desmars}(2015)}]{Desmars2015a}
{Desmars}, J. 2015, \aap, 575, A53

\bibitem[{{Desmars} {et~al.}(2015){Desmars}, {Camargo}, {Braga-Ribas},
  {Vieira-Martins}, {Assafin}, {Vachier}, {Colas}, {Ortiz}, {Duffard},
  {Morales}, {Sicardy}, {Gomes-J{\'u}nior}, \&
  {Benedetti-Rossi}}]{Desmars2015b}
{Desmars}, J., {Camargo}, J.~I.~B., {Braga-Ribas}, F., {et~al.} 2015, \aap,
  584, A96

\bibitem[{{Doressoundiram} {et~al.}(2002){Doressoundiram}, {Peixinho}, {de
  Bergh}, {Fornasier}, {Th{\'e}bault}, {Barucci}, \&
  {Veillet}}]{Doressoundiram2002}
{Doressoundiram}, A., {Peixinho}, N., {de Bergh}, C., {et~al.} 2002, The
  Astronomical Journal, 124, 2279

\bibitem[{{Dotto} {et~al.}(2003){Dotto}, {Barucci}, {Boehnhardt}, {Romon},
  {Doressoundiram}, {Peixinho}, {de Bergh}, \& {Lazzarin}}]{Dotto2003}
{Dotto}, E., {Barucci}, M.~A., {Boehnhardt}, H., {et~al.} 2003, Icarus, 162,
  408

\bibitem[{{Duffard} {et~al.}(2014{\natexlab{a}}){Duffard}, {Pinilla-Alonso},
  {Ortiz}, {Alvarez-Candal}, {Sicardy}, {Santos-Sanz}, {Morales}, {Colazo},
  {Fern{\'a}ndez-Valenzuela}, \& {Braga-Ribas}}]{Duffard2014b}
{Duffard}, R., {Pinilla-Alonso}, N., {Ortiz}, J.~L., {et~al.}
  2014{\natexlab{a}}, Astronomy and Astrophysics, 568, A79

\bibitem[{{Duffard} {et~al.}(2014{\natexlab{b}}){Duffard}, {Pinilla-Alonso},
  {Santos-Sanz}, {Vilenius}, {Ortiz}, {Mueller}, {Fornasier}, {Lellouch},
  {Mommert}, {Pal}, {Kiss}, {Mueller}, {Stansberry}, {Delsanti}, {Peixinho}, \&
  {Trilling}}]{Duffard2014}
{Duffard}, R., {Pinilla-Alonso}, N., {Santos-Sanz}, P., {et~al.}
  2014{\natexlab{b}}, Astronomy and Astrophysics, 564, A92

\bibitem[{{Fern{\'a}ndez-Valenzuela} {et~al.}(2017){Fern{\'a}ndez-Valenzuela},
  {Ortiz}, {Duffard}, {Morales}, \& {Santos-Sanz}}]{Fernandez-Valenzuela2017}
{Fern{\'a}ndez-Valenzuela}, E., {Ortiz}, J.~L., {Duffard}, R., {Morales}, N.,
  \& {Santos-Sanz}, P. 2017, \mnras, 466, 4147

\bibitem[{{Fern{\'a}ndez-Valenzuela} {et~al.}(2016){Fern{\'a}ndez-Valenzuela},
  {Ortiz}, {Duffard}, {Santos-Sanz}, \& {Morales}}]{Fernandez-Valenzuela2016}
{Fern{\'a}ndez-Valenzuela}, E., {Ortiz}, J.~L., {Duffard}, R., {Santos-Sanz},
  P., \& {Morales}, N. 2016, \mnras, 456, 2354

\bibitem[{{Gaia Collaboration} {et~al.}(2018){Gaia Collaboration}, {Brown},
  {Vallenari}, {Prusti}, {de Bruijne}, {Babusiaux}, {Bailer-Jones}, {Biermann},
  {Evans}, {Eyer}, {Jansen}, {Jordi}, {Klioner}, {Lammers}, {Lindegren},
  {Luri}, {Mignard}, {Panem}, {Pourbaix}, {Randich}, {Sartoretti}, {Siddiqui},
  {Soubiran}, {van Leeuwen}, {Walton}, {Arenou}, {Bastian}, {Cropper},
  {Drimmel}, {Katz}, {Lattanzi}, {Bakker}, {Cacciari}, {Casta{\~n}eda},
  {Chaoul}, {Cheek}, {De Angeli}, {Fabricius}, {Guerra}, {Holl}, {Masana},
  {Messineo}, {Mowlavi}, {Nienartowicz}, {Panuzzo}, {Portell}, {Riello},
  {Seabroke}, {Tanga}, {Th{\'e}venin}, {Gracia-Abril}, {Comoretto},
  {Garcia-Reinaldos}, {Teyssier}, {Altmann}, {Andrae}, {Audard},
  {Bellas-Velidis}, {Benson}, {Berthier}, {Blomme}, {Burgess}, {Busso},
  {Carry}, {Cellino}, {Clementini}, {Clotet}, {Creevey}, {Davidson}, {De
  Ridder}, {Delchambre}, {Dell'Oro}, {Ducourant},
  {Fern{\'a}ndez-Hern{\'a}ndez}, {Fouesneau}, {Fr{\'e}mat}, {Galluccio},
  {Garc{\'\i}a-Torres}, {Gonz{\'a}lez-N{\'u}{\~n}ez}, {Gonz{\'a}lez-Vidal},
  {Gosset}, {Guy}, {Halbwachs}, {Hambly}, {Harrison}, {Hern{\'a}ndez},
  {Hestroffer}, {Hodgkin}, {Hutton}, {Jasniewicz}, {Jean-Antoine-Piccolo},
  {Jordan}, {Korn}, {Krone-Martins}, {Lanzafame}, {Lebzelter}, {L{\"o}ffler},
  {Manteiga}, {Marrese}, {Mart{\'\i}n-Fleitas}, {Moitinho}, {Mora}, {Muinonen},
  {Osinde}, {Pancino}, {Pauwels}, {Petit}, {Recio-Blanco}, {Richards},
  {Rimoldini}, {Robin}, {Sarro}, {Siopis}, {Smith}, {Sozzetti}, {S{\"u}veges},
  {Torra}, {van Reeven}, {Abbas}, {Abreu Aramburu}, {Accart}, {Aerts},
  {Altavilla}, {{\'A}lvarez}, {Alvarez}, {Alves}, {Anderson}, {Andrei},
  {Anglada Varela}, {Antiche}, {Antoja}, {Arcay}, {Astraatmadja}, {Bach},
  {Baker}, {Balaguer-N{\'u}{\~n}ez}, {Balm}, {Barache}, {Barata}, {Barbato},
  {Barblan}, {Barklem}, {Barrado}, {Barros}, {Barstow}, {Bartholom{\'e}
  Mu{\~n}oz}, {Bassilana}, {Becciani}, {Bellazzini}, {Berihuete}, {Bertone},
  {Bianchi}, {Bienaym{\'e}}, {Blanco-Cuaresma}, {Boch}, {Boeche}, {Bombrun},
  {Borrachero}, {Bossini}, {Bouquillon}, {Bourda}, {Bragaglia}, {Bramante},
  {Breddels}, {Bressan}, {Brouillet}, {Br{\"u}semeister}, {Brugaletta},
  {Bucciarelli}, {Burlacu}, {Busonero}, {Butkevich}, {Buzzi}, {Caffau},
  {Cancelliere}, {Cannizzaro}, {Cantat-Gaudin}, {Carballo}, {Carlucci},
  {Carrasco}, {Casamiquela}, {Castellani}, {Castro-Ginard}, {Charlot},
  {Chemin}, {Chiavassa}, {Cocozza}, {Costigan}, {Cowell}, {Crifo}, {Crosta},
  {Crowley}, {Cuypers}, {Dafonte}, {Damerdji}, {Dapergolas}, {David}, {David},
  {de Laverny}, {De Luise}, {De March}, {de Martino}, {de Souza}, {de Torres},
  {Debosscher}, {del Pozo}, {Delbo}, {Delgado}, {Delgado}, {Di Matteo},
  {Diakite}, {Diener}, {Distefano}, {Dolding}, {Drazinos}, {Dur{\'a}n},
  {Edvardsson}, {Enke}, {Eriksson}, {Esquej}, {Eynard Bontemps}, {Fabre},
  {Fabrizio}, {Faigler}, {Falc{\~a}o}, {Farr{\`a}s Casas}, {Federici},
  {Fedorets}, {Fernique}, {Figueras}, {Filippi}, {Findeisen}, {Fonti},
  {Fraile}, {Fraser}, {Fr{\'e}zouls}, {Gai}, {Galleti}, {Garabato},
  {Garc{\'\i}a-Sedano}, {Garofalo}, {Garralda}, {Gavel}, {Gavras}, {Gerssen},
  {Geyer}, {Giacobbe}, {Gilmore}, {Girona}, {Giuffrida}, {Glass}, {Gomes},
  {Granvik}, {Gueguen}, {Guerrier}, {Guiraud}, {Guti{\'e}rrez-S{\'a}nchez},
  {Haigron}, {Hatzidimitriou}, {Hauser}, {Haywood}, {Heiter}, {Helmi}, {Heu},
  {Hilger}, {Hobbs}, {Hofmann}, {Holland}, {Huckle}, {Hypki}, {Icardi},
  {Jan{\ss}en}, {Jevardat de Fombelle}, {Jonker}, {Juh{\'a}sz}, {Julbe},
  {Karampelas}, {Kewley}, {Klar}, {Kochoska}, {Kohley}, {Kolenberg},
  {Kontizas}, {Kontizas}, {Koposov}, {Kordopatis}, {Kostrzewa-Rutkowska},
  {Koubsky}, {Lambert}, {Lanza}, {Lasne}, {Lavigne}, {Le Fustec}, {Le
  Poncin-Lafitte}, {Lebreton}, {Leccia}, {Leclerc}, {Lecoeur-Taibi},
  {Lenhardt}, {Leroux}, {Liao}, {Licata}, {Lindstr{\o}m}, {Lister}, {Livanou},
  {Lobel}, {L{\'o}pez}, {Managau}, {Mann}, {Mantelet}, {Marchal}, {Marchant},
  {Marconi}, {Marinoni}, {Marschalk{\'o}}, {Marshall}, {Martino}, {Marton},
  {Mary}, {Massari}, {Matijevi{\v{c}}}, {Mazeh}, {McMillan}, {Messina},
  {Michalik}, {Millar}, {Molina}, {Molinaro}, {Moln{\'a}r}, {Montegriffo},
  {Mor}, {Morbidelli}, {Morel}, {Morris}, {Mulone}, {Muraveva}, {Musella},
  {Nelemans}, {Nicastro}, {Noval}, {O'Mullane}, {Ord{\'e}novic},
  {Ord{\'o}{\~n}ez-Blanco}, {Osborne}, {Pagani}, {Pagano}, {Pailler},
  {Palacin}, {Palaversa}, {Panahi}, {Pawlak}, {Piersimoni}, {Pineau}, {Plachy},
  {Plum}, {Poggio}, {Poujoulet}, {Pr{\v{s}}a}, {Pulone}, {Racero}, {Ragaini},
  {Rambaux}, {Ramos-Lerate}, {Regibo}, {Reyl{\'e}}, {Riclet}, {Ripepi}, {Riva},
  {Rivard}, {Rixon}, {Roegiers}, {Roelens}, {Romero-G{\'o}mez}, {Rowell},
  {Royer}, {Ruiz-Dern}, {Sadowski}, {Sagrist{\`a} Sell{\'e}s}, {Sahlmann},
  {Salgado}, {Salguero}, {Sanna}, {Santana-Ros}, {Sarasso}, {Savietto},
  {Schultheis}, {Sciacca}, {Segol}, {Segovia}, {S{\'e}gransan}, {Shih},
  {Siltala}, {Silva}, {Smart}, {Smith}, {Solano}, {Solitro}, {Sordo}, {Soria
  Nieto}, {Souchay}, {Spagna}, {Spoto}, {Stampa}, {Steele},
  {Steidelm{\"u}ller}, {Stephenson}, {Stoev}, {Suess}, {Surdej}, {Szabados},
  {Szegedi-Elek}, {Tapiador}, {Taris}, {Tauran}, {Taylor}, {Teixeira},
  {Terrett}, {Teyssand ier}, {Thuillot}, {Titarenko}, {Torra Clotet}, {Turon},
  {Ulla}, {Utrilla}, {Uzzi}, {Vaillant}, {Valentini}, {Valette}, {van Elteren},
  {Van Hemelryck}, {van Leeuwen}, {Vaschetto}, {Vecchiato}, {Veljanoski},
  {Viala}, {Vicente}, {Vogt}, {von Essen}, {Voss}, {Votruba}, {Voutsinas},
  {Walmsley}, {Weiler}, {Wertz}, {Wevers}, {Wyrzykowski}, {Yoldas},
  {{\v{Z}}erjal}, {Ziaeepour}, {Zorec}, {Zschocke}, {Zucker}, {Zurbach}, \&
  {Zwitter}}]{Gaia-Collaboration2018}
{Gaia Collaboration}, {Brown}, A.~G.~A., {Vallenari}, A., {et~al.} 2018, \aap,
  616, A1

\bibitem[{{Gomes-J{\'u}nior} {et~al.}(2022){Gomes-J{\'u}nior}, {Morgado},
  {Benedetti-Rossi}, {Boufleur}, {Rommel}, {Banda-Huarca}, {Kilic},
  {Braga-Ribas}, \& {Sicardy}}]{Gomes-Junior2022}
{Gomes-J{\'u}nior}, A.~R., {Morgado}, B.~E., {Benedetti-Rossi}, G., {et~al.}
  2022, \mnras [\eprint[arXiv]{2201.01799}]

\bibitem[{{Horner} {et~al.}(2004){Horner}, {Evans}, \& {Bailey}}]{Horner2004a}
{Horner}, J., {Evans}, N.~W., \& {Bailey}, M.~E. 2004, MNRAS, 354, 798

\bibitem[{{Hubbard}(1984)}]{Hubbard1984}
{Hubbard}, W.~B. 1984, {Planetary interiors}

\bibitem[{{Hyodo} {et~al.}(2016){Hyodo}, {Charnoz}, {Ohtsuki}, \&
  {Genda}}]{Hyodo2016}
{Hyodo}, R., {Charnoz}, S., {Ohtsuki}, K., \& {Genda}, H. 2016, ArXiv e-prints
  [\eprint[arXiv]{1609.02396}]

\bibitem[{{Lellouch} {et~al.}(2017){Lellouch}, {Moreno}, {M{\"u}ller},
  {Fornasier}, {Santos-Sanz}, {Moullet}, {Gurwell}, {Stansberry}, {Leiva}, \&
  {Sicardy}}]{Lellouch2017}
{Lellouch}, E., {Moreno}, R., {M{\"u}ller}, T., {et~al.} 2017, \aap, 608, A45

\bibitem[{{Lomb}(1976)}]{Lomb1976}
{Lomb}, N.~R. 1976, \apss, 39, 447

\bibitem[{{Masiero} {et~al.}(2021){Masiero}, {Wright}, \&
  {Mainzer}}]{Masiero2021}
{Masiero}, J.~R., {Wright}, E.~L., \& {Mainzer}, A.~K. 2021, The Planetary
  Science Journal, 2, 32

\bibitem[{{McKinnon} {et~al.}(2020){McKinnon}, {Richardson}, {Marohnic},
  {Keane}, {Grundy}, {Hamilton}, {Nesvorn{\'y}}, {Umurhan}, {Lauer}, {Singer},
  {Stern}, {Weaver}, {Spencer}, {Buie}, {Moore}, {Kavelaars}, {Lisse}, {Mao},
  {Parker}, {Porter}, {Showalter}, {Olkin}, {Cruikshank}, {Elliott},
  {Gladstone}, {Parker}, {Verbiscer}, {Young}, \& {New Horizons Science
  Team}}]{McKinnon2020}
{McKinnon}, W.~B., {Richardson}, D.~C., {Marohnic}, J.~C., {et~al.} 2020,
  Science, 367, aay6620

\bibitem[{{Melita} {et~al.}(2017){Melita}, {Duffard}, {Ortiz}, \&
  {Campo-Bagatin}}]{Melita2017}
{Melita}, M.~D., {Duffard}, R., {Ortiz}, J.~L., \& {Campo-Bagatin}, A. 2017,
  \aap, 602, A27

\bibitem[{{Morgado} {et~al.}(2021){Morgado}, {Sicardy}, {Braga-Ribas},
  {Desmars}, {Gomes-J{\'u}nior}, {B{\'e}rard}, {Leiva}, {Ortiz},
  {Vieira-Martins}, {Benedetti-Rossi}, {Santos-Sanz}, {Camargo}, {Duffard},
  {Rommel}, {Assafin}, {Boufleur}, {Colas}, {Kretlow}, {Beisker}, {Sfair},
  {Snodgrass}, {Morales}, {Fern{\'a}ndez-Valenzuela}, {Amaral}, {Amarante},
  {Artola}, {Backes}, {Bath}, {Bouley}, {Buie}, {Cacella}, {Colazo}, {Colque},
  {Dauvergne}, {Dominik}, {Emilio}, {Erickson}, {Evans}, {Fabrega-Polleri},
  {Garcia-Lambas}, {Giacchini}, {Hanna}, {Herald}, {Hesler}, {Hinse},
  {Jacques}, {Jehin}, {J{\o}rgensen}, {Kerr}, {Kouprianov}, {Levine}, {Linder},
  {Maley}, {Machado}, {Maquet}, {Maury}, {Melia}, {Meza}, {Mondon}, {Moura},
  {Newman}, {Payet}, {Pereira}, {Pollock}, {Poltronieri}, {Quispe-Huaynasi},
  {Reichart}, {de Santana}, {Schneiter}, {Sieyra}, {Skottfelt}, {Soulier},
  {Starck}, {Thierry}, {Torres}, {Trabuco}, {Unda-Sanzana}, {Yamashita},
  {Winter}, {Zapata}, \& {Zuluaga}}]{Morgado2021}
{Morgado}, B.~E., {Sicardy}, B., {Braga-Ribas}, F., {et~al.} 2021, \aap, 652,
  A141

\bibitem[{{M{\"u}ller} {et~al.}(2020){M{\"u}ller}, {Lellouch}, \&
  {Fornasier}}]{Mueller2020}
{M{\"u}ller}, T., {Lellouch}, E., \& {Fornasier}, S. 2020, {Trans-Neptunian
  objects and Centaurs at thermal wavelengths}, ed. D.~{Prialnik}, M.~A.
  {Barucci}, \& L.~{Young}, 153--181

\bibitem[{{Ortiz} {et~al.}(2015){Ortiz}, {Duffard}, {Pinilla-Alonso},
  {Alvarez-Candal}, {Santos-Sanz}, {Morales}, {Fern{\'a}ndez-Valenzuela},
  {Licandro}, {Campo Bagatin}, \& {Thirouin}}]{Ortiz2015}
{Ortiz}, J.~L., {Duffard}, R., {Pinilla-Alonso}, N., {et~al.} 2015, Astronomy
  and Astrophysics, 576, A18

\bibitem[{{Ortiz} {et~al.}(2003){Ortiz}, {Guti{\'e}rrez}, {Casanova}, \&
  {Sota}}]{Ortiz2003}
{Ortiz}, J.~L., {Guti{\'e}rrez}, P.~J., {Casanova}, V., \& {Sota}, A. 2003,
  Astronomy and Astrophysics, 407, 1149

\bibitem[{{Ortiz} {et~al.}(2020{\natexlab{a}}){Ortiz}, {Kiss}, {Santos-Sanz},
  {Mueller}, {Duffard}, {Sicardy}, {Braga-Ribas}, {Benedetti-Rossi}, {Morales},
  {Desmars}, {Lecacheux}, {Vieira-Martins}, {Camargo}, {Assafin}, {Colas},
  {Fern\'andez-Valenzuela}, {Vara-Lubiano}, {Guti\'errez}, {Alvarez-Candal},
  {Morgado}, \& {Rommel}}]{Ortiz2020abstract}
{Ortiz}, J.~L., {Kiss}, C., {Santos-Sanz}, P., {et~al.} 2020{\natexlab{a}}, in
  European Planetary Science Congress, EPSC2020--686

\bibitem[{{Ortiz} {et~al.}(2017){Ortiz}, {Santos-Sanz}, {Sicardy},
  {Benedetti-Rossi}, {B{\'e}rard}, {Morales}, {Duffard}, {Braga-Ribas}, {Hopp},
  {Ries}, {Nascimbeni}, {Marzari}, {Granata}, {P{\'a}l}, {Kiss}, {Pribulla},
  {Kom{\v z}{\'{\i}}k}, {Hornoch}, {Pravec}, {Bacci}, {Maestripieri}, {Nerli},
  {Mazzei}, {Bachini}, {Martinelli}, {Succi}, {Ciabattari}, {Mikuz},
  {Carbognani}, {Gaehrken}, {Mottola}, {Hellmich}, {Rommel},
  {Fern{\'a}ndez-Valenzuela}, {Campo Bagatin}, {Cikota}, {Cikota}, {Lecacheux},
  {Vieira-Martins}, {Camargo}, {Assafin}, {Colas}, {Behrend}, {Desmars},
  {Meza}, {Alvarez-Candal}, {Beisker}, {Gomes-Junior}, {Morgado}, {Roques},
  {Vachier}, {Berthier}, {Mueller}, {Madiedo}, {Unsalan}, {Sonbas}, {Karaman},
  {Erece}, {Koseoglu}, {Ozisik}, {Kalkan}, {Guney}, {Niaei}, {Satir},
  {Yesilyaprak}, {Puskullu}, {Kabas}, {Demircan}, {Alikakos}, {Charmandaris},
  {Leto}, {Ohlert}, {Christille}, {Szak{\'a}ts}, {Tak{\'a}csn{\'e} Farkas},
  {Varga-Vereb{\'e}lyi}, {Marton}, {Marciniak}, {Bartczak}, {Santana-Ros},
  {Butkiewicz-B{\c a}k}, {Dudzi{\'n}ski}, {Al{\'{\i}}-Lagoa}, {Gazeas},
  {Tzouganatos}, {Paschalis}, {Tsamis}, {S{\'a}nchez-Lavega},
  {P{\'e}rez-Hoyos}, {Hueso}, {Guirado}, {Peris}, \&
  {Iglesias-Marzoa}}]{Ortiz2017}
{Ortiz}, J.~L., {Santos-Sanz}, P., {Sicardy}, B., {et~al.} 2017, \nat, 550, 219

\bibitem[{{Ortiz} {et~al.}(2020{\natexlab{b}}){Ortiz}, {Santos-Sanz},
  {Sicardy}, {Benedetti-Rossi}, {Duffard}, {Morales}, {Braga-Ribas},
  {Fern{\'a}ndez-Valenzuela}, {Nascimbeni}, {Nardiello}, {Carbognani}, {Buzzi},
  {Aletti}, {Bacci}, {Maestripieri}, {Mazzei}, {Mikuz}, {Skvarc}, {Ciabattari},
  {Lavalade}, {Scarfi}, {Mari}, {Conjat}, {Sposetti}, {Bachini}, {Succi},
  {Mancini}, {Alighieri}, {Dal Canto}, {Masucci}, {Vara-Lubiano},
  {Guti{\'e}rrez}, {Desmars}, {Lecacheux}, {Vieira-Martins}, {Camargo},
  {Assafin}, {Colas}, {Beisker}, {Behrend}, {Mueller}, {Meza}, {Gomes-Junior},
  {Roques}, {Vachier}, {Mottola}, {Hellmich}, {Campo Bagatin},
  {Alvarez-Candal}, {Cikota}, {Cikota}, {Christille}, {P{\'a}l}, {Kiss},
  {Pribulla}, {Kom{\v{z}}{\'\i}k}, {Madiedo}, {Charmandaris}, {Alikakos},
  {Szak{\'a}ts}, {Farkas-Tak{\'a}cs}, {Varga-Vereb{\'e}lyi}, {Marton},
  {Marciniak}, {Bartczak}, {Butkiewicz-Ba{\c{k}}}, {Dudzi{\'n}ski},
  {Al{\'\i}-Lagoa}, {Gazeas}, {Paschalis}, {Tsamis}, {Guirado}, {Peris},
  {Iglesias-Marzoa}, {Schnabel}, {Manzano}, {Navarro}, {Perell{\'o}},
  {Vecchione}, {Noschese}, \& {Morrone}}]{Ortiz2020}
{Ortiz}, J.~L., {Santos-Sanz}, P., {Sicardy}, B., {et~al.} 2020{\natexlab{b}},
  \aap, 639, A134

\bibitem[{{Ortiz} {et~al.}(2012){Ortiz}, {Thirouin}, {Campo Bagatin},
  {Duffard}, {Licandro}, {Richardson}, {Santos-Sanz}, {Morales}, \&
  {Benavidez}}]{Ortiz2012b}
{Ortiz}, J.~L., {Thirouin}, A., {Campo Bagatin}, A., {et~al.} 2012, \mnras,
  419, 2315

\bibitem[{{Peixinho} {et~al.}(2012){Peixinho}, {Delsanti}, {Guilbert-Lepoutre},
  {Gafeira}, \& {Lacerda}}]{Peixinho2012}
{Peixinho}, N., {Delsanti}, A., {Guilbert-Lepoutre}, A., {Gafeira}, R., \&
  {Lacerda}, P. 2012, Astronomy and Astrophysics, 546, A86

\bibitem[{{Press} {et~al.}(1992){Press}, {Teukolsky}, {Vetterling}, \&
  {Flannery}}]{Press1992}
{Press}, W.~H., {Teukolsky}, S.~A., {Vetterling}, W.~T., \& {Flannery}, B.~P.
  1992, {Numerical recipes in FORTRAN. The art of scientific computing}
  (Cambridge: University Press)

\bibitem[{{Proudfoot} \& {Ragozzine}(2019)}]{Proudfoot2019}
{Proudfoot}, B. C.~N. \& {Ragozzine}, D. 2019, \aj, 157, 230

\bibitem[{{Rabinowitz} {et~al.}(2007){Rabinowitz}, {Schaefer}, \&
  {Tourtellotte}}]{Rabinowitz2007}
{Rabinowitz}, D.~L., {Schaefer}, B.~E., \& {Tourtellotte}, S.~W. 2007, The
  Astronomical Journal, 133, 26

\bibitem[{{Ruprecht} {et~al.}(2015){Ruprecht}, {Bosh}, {Person}, {Bianco},
  {Fulton}, {Gulbis}, {Bus}, \& {Zangari}}]{Ruprecht2015}
{Ruprecht}, J.~D., {Bosh}, A.~S., {Person}, M.~J., {et~al.} 2015, \icarus, 252,
  271

\bibitem[{{Santos-Sanz} {et~al.}(2022){Santos-Sanz}, {Ortiz}, {Sicardy},
  {Popescu}, {Benedetti-Rossi}, {Morales}, {Vara-Lubiano}, {Camargo},
  {Pereira}, {Rommel}, {Assafin}, {Desmars}, {Braga-Ribas}, {Duffard}, {Marques
  Oliveira}, {Vieira-Martins}, {Fern{\'a}ndez-Valenzuela}, {Morgado}, {Acar},
  {Anghel}, {Atalay}, {Ate{\c{s}}}, {Baki{\c{s}}}, {Bakis}, {Eker}, {Erece},
  {Kaspi}, {Kayhan}, {Kilic}, {Kilic}, {Manulis}, {Nedelcu}, {Niaei}, {Nir},
  {Ofek}, {Ozisik}, {Petrescu}, {Satir}, {Solmaz}, {Sonka}, {Tekes}, {Unsalan},
  {Yesilyaprak}, {Anghel}, {Berte{\c{s}}teanu}, {Curelaru}, {Danescu},
  {Dumitrescu}, {Gherase}, {Hudin}, {Stoian}, {Tercu}, {Truta}, {Turcu},
  {Vantdevara}, {Belskaya}, {Dementiev}, {Gazeas}, {Karampotsiou}, {Kashuba},
  {Kiss}, {Koshkin}, {Kozhukhov}, {Krugly}, {Lecacheux}, {Pal},
  {P{\"u}sk{\"u}ll{\"u}}, {Szakats}, {Zhukov}, {Bamberger}, {Mondon},
  {Perell{\'o}}, {Pratt}, {Schnabel}, {Selva}, {Teng}, {Tigani}, {Tsamis},
  {Weber}, {Wells}, {Kalkan}, {Kudak}, {Marciniak}, {Ogloza}, {{\"O}zdemir},
  {Pak{\v{s}}tiene}, {Perig}, \& {{\.Z}ejmo}}]{Santos-Sanz2022}
{Santos-Sanz}, P., {Ortiz}, J.~L., {Sicardy}, B., {et~al.} 2022, \aap, 664,
  A130

\bibitem[{{Sicardy} {et~al.}(2019){Sicardy}, {Leiva}, {Renner}, {Roques}, {El
  Moutamid}, {Santos-Sanz}, \& {Desmars}}]{Sicardy2019}
{Sicardy}, B., {Leiva}, R., {Renner}, S., {et~al.} 2019, Nature Astronomy, 3,
  146

\bibitem[{{Sicardy} {et~al.}(2011){Sicardy}, {Ortiz}, {Assafin}, {Jehin},
  {Maury}, {Lellouch}, {Hutton}, {Braga-Ribas}, {Colas}, {Hestroffer},
  {Lecacheux}, {Roques}, {Santos-Sanz}, {Widemann}, {Morales}, {Duffard},
  {Thirouin}, {Castro-Tirado}, {Jel{\'{\i}}nek}, {Kub{\'a}nek}, {Sota},
  {S{\'a}nchez-Ram{\'{\i}}rez}, {Andrei}, {Camargo}, {da Silva Neto}, {Gomes},
  {Martins}, {Gillon}, {Manfroid}, {Tozzi}, {Harlingten}, {Saravia}, {Behrend},
  {Mottola}, {Melendo}, {Peris}, {Fabregat}, {Madiedo}, {Cuesta}, {Eibe},
  {Ull{\'a}n}, {Organero}, {Pastor}, {de Los Reyes}, {Pedraz}, {Castro}, {de La
  Cueva}, {Muler}, {Steele}, {Cebri{\'a}n},
  {Monta{\~n}{\'e}s-Rodr{\'{\i}}guez}, {Oscoz}, {Weaver}, {Jacques}, {Corradi},
  {Santos}, {Reis}, {Milone}, {Emilio}, {Guti{\'e}rrez}, {V{\'a}zquez}, \&
  {Hern{\'a}ndez-Toledo}}]{Sicardy2011}
{Sicardy}, B., {Ortiz}, J.~L., {Assafin}, M., {et~al.} 2011, \nat, 478, 493

\bibitem[{{Sicardy} {et~al.}(2020){Sicardy}, {Renner}, {Leiva}, {Roques}, {El
  Moutamid}, {Santos-Sanz}, \& {Desmars}}]{Sicardy2020}
{Sicardy}, B., {Renner}, S., {Leiva}, R., {et~al.} 2020, {The dynamics of rings
  around Centaurs and Trans-Neptunian objects}, ed. D.~{Prialnik}, M.~A.
  {Barucci}, \& L.~{Young}, 249--269

\bibitem[{{Sickafoose} {et~al.}(2020){Sickafoose}, {Bosh}, {Emery}, {Person},
  {Zuluaga}, {Womack}, {Ruprecht}, {Bianco}, \& {Zangari}}]{Sickafoose2020}
{Sickafoose}, A.~A., {Bosh}, A.~S., {Emery}, J.~P., {et~al.} 2020, \mnras, 491,
  3643

\bibitem[{{Snedecor} \& {Cochran}(1989)}]{Snedecor1989}
{Snedecor}, G.~W. \& {Cochran}, W.~G. 1989, Statistical Methods, Eighth
  Edition, Iowa State University Press

\bibitem[{{Souami} {et~al.}(2020){Souami}, {Braga-Ribas}, {Sicardy}, {Morgado},
  {Ortiz}, {Desmars}, {Camargo}, {Vachier}, {Berthier}, {Carry}, {Anderson},
  {Showers}, {Thomason}, {Maley}, {Thomas}, {Buie}, {Leiva}, {Keller},
  {Vieira-Martins}, {Assafin}, {Santos-Sanz}, {Morales}, {Duffard},
  {Benedetti-Rossi}, {Gomes-J{\'u}nior}, {Boufleur}, {Pereira}, {Margoti},
  {Pavlov}, {George}, {Oesper}, {Bardecker}, {Dunford}, {Kehrli}, {Spencer},
  {Cota}, {Garcia}, {Lara}, {McCandless}, {Self}, {Lecacheux}, {Frappa},
  {Dunham}, \& {Emilio}}]{Souami2020}
{Souami}, D., {Braga-Ribas}, F., {Sicardy}, B., {et~al.} 2020, \aap, 643, A125

\bibitem[{{Stellingwerf}(1978)}]{Stellingwerf1978a}
{Stellingwerf}, R.~F. 1978, The Astrophysical Journal, 224, 953

\bibitem[{{Stetson}(1987)}]{Stetson1987}
{Stetson}, P.~B. 1987, \pasp, 99, 191

\bibitem[{{Th{\'e}bault} \& {Doressoundiram}(2003)}]{Thebault2003}
{Th{\'e}bault}, P. \& {Doressoundiram}, A. 2003, \icarus, 162, 27

\bibitem[{{Thirouin} {et~al.}(2014){Thirouin}, {Noll}, {Ortiz}, \&
  {Morales}}]{Thirouin2014}
{Thirouin}, A., {Noll}, K.~S., {Ortiz}, J.~L., \& {Morales}, N. 2014, \aap,
  569, A3

\bibitem[{{van Belle}(1999)}]{vanBelle1999}
{van Belle}, G.~T. 1999, \pasp, 111, 1515

\bibitem[{{Vara-Lubiano} {et~al.}(2022){Vara-Lubiano}, {Benedetti-Rossi},
  {Santos-Sanz}, {Ortiz}, {Sicardy}, {Popescu}, {Morales}, {Rommel}, {Morgado},
  {Pereira}, {{\'A}lvarez-Candal}, {Fern{\'a}ndez-Valenzuela}, {Souami},
  {Ilic}, {Vince}, {Bachev}, {Semkov}, {Nedelcu}, {{\c{S}}onka}, {Hudin},
  {Boaca}, {Inceu}, {Curelaru}, {Gherase}, {Turcu}, {Moldovan}, {Mircea},
  {Predatu}, {Teodorescu}, {Stoian}, {Juravle}, {Braga-Ribas}, {Desmars},
  {Duffard}, {Lecacheux}, {Camargo}, {Assafin}, {Vieira-Martins}, {Pribulla},
  {Hus{\'a}rik}, {Sivani{\v{c}}}, {Pal}, {Szakats}, {Kiss}, {Alonso-Santiago},
  {Frasca}, {Szab{\'o}}, {Derekas}, {Szigeti}, {Drozdz}, {Ogloza},
  {Skvar{\v{c}}}, {Ciabattari}, {Delincak}, {Di Marcantonio}, {Iafrate},
  {Coretti}, {Baldini}, {Baruffetti}, {Kl{\"o}s}, {Dumitrescu}, {Miku{\v{z}}},
  \& {Mohar}}]{Vara-Lubiano2022}
{Vara-Lubiano}, M., {Benedetti-Rossi}, G., {Santos-Sanz}, P., {et~al.} 2022,
  \aap, 663, A121

\bibitem[{{Zacharias} {et~al.}(2012){Zacharias}, {Finch}, {Girard}, {Henden},
  {Bartlett}, {Monet}, \& {Zacharias}}]{Zacharias2012}
{Zacharias}, N., {Finch}, C.~T., {Girard}, T.~M., {et~al.} 2012, VizieR Online
  Data Catalog, I/322A

\end{thebibliography}

\appendix

\section{Monte Carlo distributions}
\label{app:probability_density_functions}

\begin{figure*}
    \centering
    \includegraphics[width=17cm]{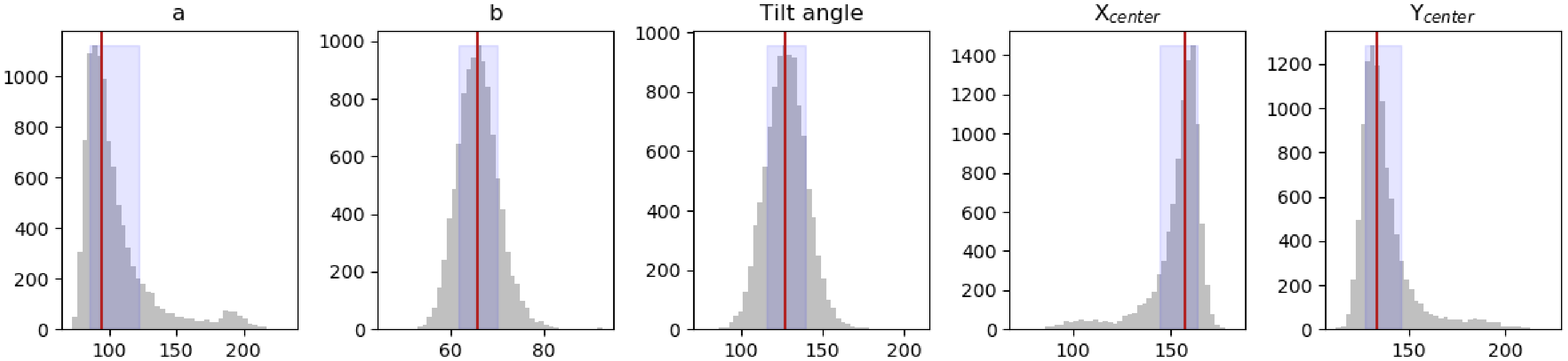}
    \caption{Monte Carlo distributions for S1 of the semi-axes $a$ and $b$ (in km), the tilt angle (in degrees) and the coordinates $(x,y)$ of the center of the ellipse. The vertical red lines show the value of the parameters for the best elliptical fit via least-squares minimization, and the blue shaded area delimits the 67\% confidence interval, see Table \ref{tab:results_from_fits}.}
    \label{fig:S1_MCD}
\end{figure*}

\begin{figure*}
    \centering
    \includegraphics[width=17cm]{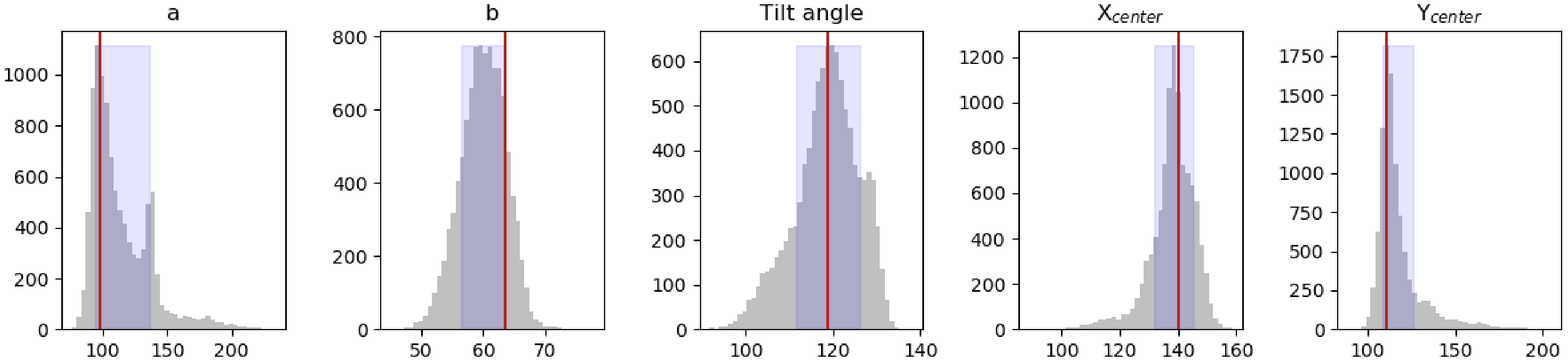}
    \caption{Monte Carlo distributions for S2 of the semi-axes $a$ and $b$ (in km), the tilt angle (in degrees) and the coordinates $(x,y)$ of the center of the ellipse. The vertical red lines show the value of the parameters for the best elliptical fit via least-squares minimization, and the blue shaded area delimits the 67\% confidence interval, see Table \ref{tab:results_from_fits}.}
    \label{fig:S2_MCD}
\end{figure*}

\begin{figure*}
    \centering
    \includegraphics[width=17cm]{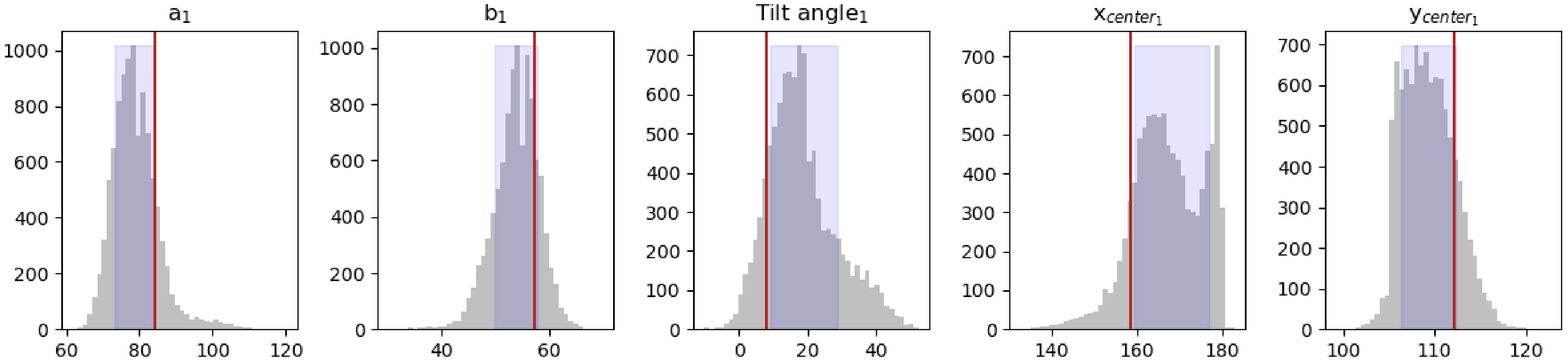}
    \caption{Monte Carlo distributions for S3, ellipse 1, of the semi-axes $a$ and $b$ (in km), the tilt angle (in degrees) and the coordinates $(x,y)$ of the center of the ellipse. The vertical red lines show the value of the parameters for the best elliptical fit via least-squares minimization, and the blue shaded area delimits the 67\% confidence interval, see Table \ref{tab:results_from_fits}.}
    \label{fig:S3_E1_MCD}
\end{figure*}

\begin{figure*}
    \centering
    \includegraphics[width=17cm]{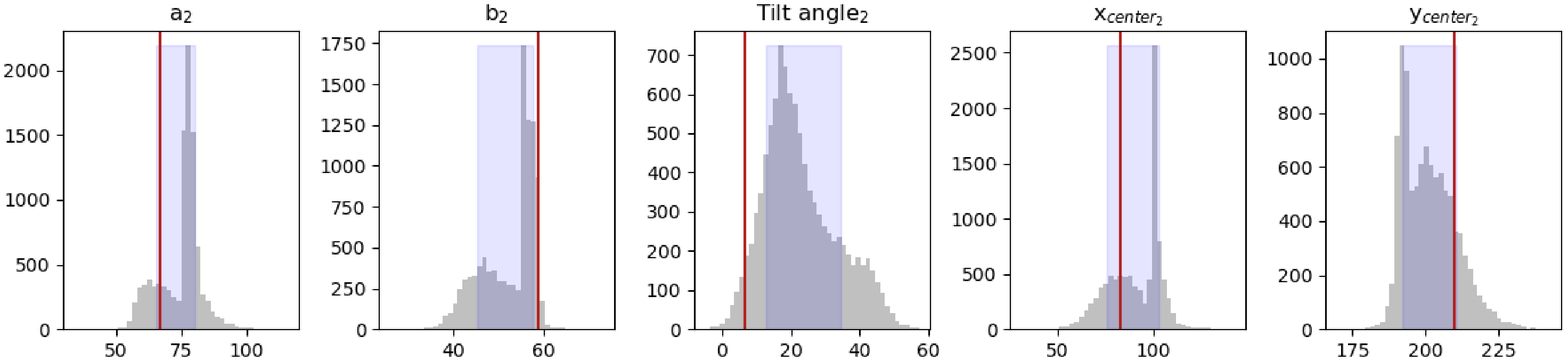}
    \caption{Monte Carlo distributions for S3, ellipse 2, of the semi-axes $a$ and $b$ (in km), the tilt angle (in degrees) and the coordinates $(x,y)$ of the center of the ellipse. The vertical red lines show the value of the parameters for the best elliptical fit via least-squares minimization, and the blue shaded area delimits the 67\% confidence interval, see Table \ref{tab:results_from_fits}.}
    \label{fig:S3_E2_MCD}
\end{figure*}

Monte Carlo distributions of the different parameters that define the elliptical fits to the chords obtained through the Monte Carlo method for the different scenarios considered in Section \ref{sec:analysis} as follows: S1 in which Manzanare's chord is shifted (Figure \ref{fig:S1_MCD}), S2 where all chords are aligned (Figure \ref{fig:S2_MCD}), S3 in which a binary object is considered (see Figure \ref{fig:S3_E1_MCD} and \ref{fig:S3_E2_MCD}, for the ellipses 1 and 2, respectively).

\end{document}